\documentclass[12pt,preprint]{aastex}
\usepackage{amsmath}
\newcommand\cm{{\,\rm cm}}
\newcommand\pcc{{\,\rm cm}^{-3}}
\newcommand\K{{\;\rm K}}
\newcommand\cs{c_s}

\newcommand\yr{{\;\rm yr}}

\newcommand\Msun{{\;\rm\,M_\odot}}

\newcommand\kms{{\;\rm km\; s^{-1}}}

\newcommand\pc{{\;\rm\,pc}}

\newcommand\simgt{\lower.5ex\hbox{$\; \buildrel > \over \sim \;$}}
\newcommand\simlt{\lower.5ex\hbox{$\; \buildrel < \over \sim \;$}}

\begin{document}
\title{Dense core formation in supersonic turbulent converging flows}
\author{Hao Gong \& Eve C.\ Ostriker}

\affil{Department of Astronomy, University of Maryland, College Park, MD 20742-2421}
\email{hgong@astro.umd.edu, ostriker@astro.umd.edu}

\begin{abstract}
We use numerical hydrodynamic simulations to investigate prestellar
core formation in the dynamic environment of giant molecular clouds
(GMCs), focusing on planar post-shock layers produced by colliding
turbulent flows. A key goal is to test how core evolution and
properties depend on the velocity dispersion in the parent cloud; our
simulation suite consists of 180 models with inflow Mach numbers
${\cal M}\equiv v/c_s =1.1-9$. At all Mach numbers, our models show
that turbulence and self-gravity collect gas within post-shock regions into
filaments at the same time as overdense areas within these filaments
condense into cores.  This morphology, together with the subsonic
velocities we find inside cores, is similar to observations.  We
extend previous results showing that core collapse develops in an
``outside-in'' manner, with density and velocity approaching the
Larson-Penston asymptotic solution.  The time for the first core to
collapse depends on Mach number as $t_{\rm coll} \propto {\cal
M}^{-1/2}\rho_0^{-1/2}$, for $\rho_0$ the mean pre-shock density, 
consistent with analytic estimates.  Core building takes
10 times as long as core collapse, which lasts a few $\times 10^5$
yrs, consistent with observed prestellar core lifetimes.  Core shapes
change from oblate to prolate as they evolve.  To define cores, we use
isosurfaces of the gravitational potential.  We compare to cores
defined using the potential computed from projected surface density,
finding good agreement for core masses and sizes; this offers a new
way to identify cores in observed maps.  Cores with masses varying by
three orders of magnitude ($\sim 0.05 - 50 M_\odot$) are identified
in our high-$\cal M$ simulations, with a much smaller mass range for
models having low $\cal M$.  We halt each simulation when the first
core collapses; at that point, only the more massive cores in each
model are gravitationally bound, with $E_{\rm th} + E_g <0$. Stability
analysis of post-shock layers predicts that the first core to collapse
will have mass $M \propto v^{-1/2} \rho_0^{-1/2} T^{7/4}$, and that
the minimum mass for cores formed at late times will have $M\propto
v^{-1} \rho_0^{-1/2} T^2$, for $T$ the temperature.  From our 
simulations, the median mass lies between these two relations.  At 
the time we halt the simulations, the $M$ vs. $v$ relation is shallower 
for bound cores than unbound cores; with further evolution the small 
cores may evolve to become bound, steeping the $M$ vs. $v$ relation.
\end{abstract}


\section{Introduction}

Star formation begins with the creation of dense molecular cores, and
understanding how cores grow and evolve is essential to identifying
the origin of stellar properties \citep{shu87,mcke07,andr08}. Through
the 1990s, the prevailing theoretical picture was of slow core
formation and evolution mediated by ambipolar diffusion, followed by
core collapse initiated from a quasistatic, centrally-concentrated
state \citep[e.g.,][]{mous87,mous99}. Current observations, however,
indicate that magnetic field strengths are insufficient to provide the
dominant support of molecular cores \citep{trol08}.  In addition, over
the past decade, a conception of star formation has emerged in which
supersonic turbulence drives structure and evolution within giant
molecular clouds (GMCs) on a wide range of scales
\citep[e.g.,][]{ball07, mcke07}. Because supersonic turbulence can
compress gas to densities at which gravitational collapse can rapidly
occur, it is likely to be important in the initiation of prestellar
cores. Ultimately, models of core formation and evolution must take
into account both moderate magnetic fields (with diffusion) and strong
turbulence \citep{kudo08,naka08}.  In order to gain insight into the
physics involved, however, it is informative to focus on individual
limiting cases and explore dependence on parameters.  Here, following
\citet{gong09} but generalizing to three dimensions, we consider core
building and evolution in the turbulence-dominated, unmagnetized
limit.

Observations of dense cores in GMCs have provided detailed information
on individual core properties as well as statistics of core
populations \citep[see e.g., the reviews
of][]{difr07,ward07,berg07,andr08}. These properties, including
internal structure and kinematics, durations of different evolutionary
stages, and distribution of core masses, constrain core formation
theories.  In terms of structure, cores are observed to be centrally
concentrated at all stages, with the specific profile fits differing
depending on the stage of evolution.  Cores can generally be fit with
a uniform-density inner region surrounded by a power law $\propto
r^{-2}$ \citep[e.g.,][]{shir00,bacm00, alve01,kand05,kirk05}; this
shape is consistent with expectations for both static Bonnor-Ebert
(BE) pressure-supported isothermal equilibria \citep{bonn56,eber55},
and for collapsing isothermal spheres \citep{
bode68,lars69,pens69}. The center-to-edge density contrast is
frequently larger than the maximum possible for a stable BE sphere,
however, and the inferred temperatures based on static BE fits are
also often larger than observed temperatures. Although in principle
some support could be provided by magnetic fields
\citep[e.g.,][]{ciol94}, another possibility is that these
``supercritical'' cores are in fact collapsing rather than static
\citep{dapp09,gong09}.

In terms of kinematics, dense, low-mass cores generally have subsonic
internal velocity dispersions, whether for isolated cores or for cores
found in clusters \citep[e.g.,][]{myer83,good98,case02,tafa04,
kirk07,andr07,lada08}. Some prestellar cores also show indications of
subsonic inward motions throughout their interiors based on asymmetry
of molecular lines that trace dense gas
\citep[e.g.,][]{lee99,lee01,sohn07}.  For cores containing protostars,
signatures of supersonic inward motions on small scales ($\sim 0.01 -
0.1 \pc$) have been observed \citep[e.g.,][]{greg97,difr01}; these are
believed to be indicative of gravitationally-induced infall. In very
recent work, \citet{pine10} have used $\mathrm{NH}_3$ observations to
identify a sharp transition from supersonic to subsonic velocity
dispersion from outer to inner regions in the core B5 in Perseus.

Several recent statistical studies have reached similar conclusions
regarding the durations of successive stages of core evolution
\citep[e.g.,][]{ward07, enoc08,evan09}, with prestellar and
protostellar (class 0) stages having comparable lifetimes. The typical
duration for each of these stages is a few times the gravitational
free-fall time
\begin{equation}\label{t_ff}
t_{ff} = \left(\frac{3 \pi}{32 G \bar{\rho}} \right)^{1/2} 
= 4.3 \times 10^5 \yr \left(\frac{\bar{n}_H}{10^4 \pcc}\right)^{-1/2}
\end{equation}
at the mean core density $\bar{\rho} = 1.4 m_H \bar{n}_H$, amounting
to $\sim$ 1 -- 5 $\times 10^5$ yr for typical conditions. With
prestellar lifetimes considerably below the ambipolar diffusion time
for strong magnetic field $t_{AD} \approx 10 t_{ff}$
\citep[e.g.][]{mous99}, this suggests that observed cores are
trans-critical or supercritical \citep[see][]{ciol01} with respect to
the magnetic field.\footnote{ The critical mass-to-magnetic-flux
defines the minimum that permits gravitational collapse in the
field-freezing limit \citep[e.g.][]{mest56,mous76,naka78}.}  This
conclusion is also supported by magnetic field Zeeman observations
\citep{trol08}, indicating that cores have mean mass-to-magnetic-flux
ratios twice the critical value. Thus, magnetic field effects appear
to be sub-dominant in terms of supporting cores against collapse, and
ambipolar diffusion does not appear to control the dynamics of core
formation and evolution. As magnetic fields are non-negligible,
however, magnetohydrodynamic (MHD) stresses may still affect GMC and
core dynamics.

Empirical measurements of core mass functions (CMFs)
\citep[e.g.,][]{mott98,test98,john00,
john01,mott01,onis02,beut04,reid05,reid06,stan06,enoc06,alve07,iked07,iked09b,
iked09a,nutt07,simp08,kony10} show that CMFs have a remarkable
similarity in shape to stellar initial mass functions (IMFs, see
e.g. \citealt{krou01, chab05}), with a shift toward lower mass by a
factor of 3 -- 4 \citep[see e.g.,][]{alve07,rath09}.  The
characteristic/turnover mass of observed CMFs ranges from 0.1 -- 3
$\Msun$, although there are uncertainties in this associated with lack 
of spatial resolution at the low mass end.

Many theoretical efforts have contributed to interpreting the observed
properties of cores.  The classic work of \citet{bonn56} and
\citet{eber55} provided the foundation of later studies, by
determining the maximum mass of a static isothermal sphere that is
dynamically stable.  In terms of the boundary pressure
$P_\mathrm{edge} = \rho_\mathrm{edge} c_s^2$ or mean internal density
$\bar{\rho} = 2.5 \rho_\mathrm{edge}$, this maximum stable mass is
\begin{equation}\label{m_be}
M_{BE} = 1.2 \frac{c_s^4}{(G^3 P_{\mathrm{edge}})^{1/2}} 
= 1.9 \frac{c_s^3}{(G^3\bar{\rho})^{1/2}} 
= 2.3\Msun \left(\frac{\bar{n}_H}{10^4 \pcc} \right)^{-1/2} \left(\frac{T}{10K}\right)^{3/2}.
\end{equation}
Here, $c_s = (kT/\mu)^{1/2}$ is the internal sound speed in the core.

Over many years, numerical simulations have been used to investigate
isothermal collapse of individual, pre-existing cores
\citep{bode68,lars69,pens69,hunt77, fost93,
ogin99,henn03,moto03,voro05,gome07,burk09}. These simulations include
initiation from static configurations that are unstable, and
initiation from static, stable configurations that are subjected to
imposed compression, either from enhanced external pressure or a
converging velocity field, or a core-core collision.  A common feature
of the results is that the collapse generally starts from outside and
propagates in as the central density increases. At the time of
singularity formation, the density profile approaches the
``Larson-Penston'' asymptotic solution $\rho = 8.86 c_s^2/(4 \pi G
r^2)$ and the central velocity is comparable to the value $-3.28 c_s$
derived by \citet{lars69} and \citet{pens69}.  However, these previous
studies have not considered core evolution within the larger context,
in particular including the process of \emph{core formation}.  Since
the formation process may affect later evolution, it is important to
develop unified models.

At GMC scales, a number of groups have investigated the CMFs that
result from numerical simulations of turbulent, self-gravitating
systems \citep[see
e.g.,][]{kles01,gamm03,bonn03,li04,till04,heit08,clar08,offn08,basu09,smit09}.
These models have shown -- for certain parts of parameter space --
features that are in accord with observed CMFs: mass functions
dominated by the low end with a peak and turnover near $1 \Msun$, and
a high-mass power-law slope (at least marginally) consistent with the
Salpeter value. These simulations have not, however, had sufficient
resolution to investigate the internal properties of individual cores
that form. In addition, these studies have not quantified how the core
masses depend on the large-scale properties of the turbulent medium
(see below).

Taking the previous numerical simulations of individual cores one step
further, \citet{gong09} initiated a study of dynamically induced core
formation and evolution in supersonic converging flows, focusing on
the spherical case.  In these simulations, the density is initially
uniform everywhere: no initial core structure is assumed. Instead,
dense cores form inside a spherical shock that propagates outward
within the converging flow. Over time, cores become increasingly
stratified as their masses grow. Eventually, the core collapses to
create a protostar following the same ``outside-in'' pattern as in
models initiated from static conditions. Subsequently, the dense
envelope falls into the center via an inside-out rarefaction wave
\citep{shu77,hunt77}; this is followed by a stage of late accretion if
the converging flow on large scales continues to be maintained. The
unified formation and evolution model of \citet{gong09} explains many
observed core properties, including BE-sphere-like density profiles,
subsonic internal velocities within cores, and short core lifetimes
with comparable prestellar and protostellar durations.  \citet{gong09}
also found that the inflow velocity of the converging flows affects
core lifetimes, masses, sizes and accretion histories.  Realistic
supersonic inflows in clouds are not spherical, however, while mass
inflow rates are affected by geometry. Thus, the quantitative results
for masses, lifetimes, etc., as a function of Mach number and ambient
density may differ for more realistic geometry.

Numerical results on core formation have not reached consensus on how the characteristic 
mass in the CMF, $M_c$, depends on the bulk properties of the cloud -- its
mean density $\rho_0 = \langle \rho \rangle$, sound speed $c_s$, and turbulent
velocity dispersion $v_{\mathrm{turb}}$. Some have suggested that the 
Jeans mass of the cloud at its mean density ($M_J = c_s^3 \pi^{3/2} (G^3 \rho_0)^{-1/2}$) 
determines $M_c$ in the CMF (e.g,. \citealt{kles01,bonn06}), while
others have found values of $M_c$ well below $M_J$ 
\citep[see e.g.,][]{gamm03,li04}. As noted by \citet{mcke07}, the difference
between these conclusions is likely related to the Mach number of turbulence:
the value found for $M_c/M_J$ is lower in simulations where the Mach
number $\mathcal{M} \equiv v_{\mathrm{turb}}/c_s$ is higher. Indeed, more recent
simulations by \citet{clar08} provide some indication that increasing $\mathcal{M}$ lowers
the value of $M_c$ in the CMF; they did not, however, conduct a full 
parameter study.

Supersonic turbulence makes the density in a GMC highly non-uniform,
creating a log-normal probability distribution function (PDF) in which
most of the volume is at densities below $\rho_0$ and most of the mass
is at densities above $\rho_0$
\citep[e.g.,][]{vazq94,pado97,ostr99}. Given that the log-normal PDF
allows for a range of Jeans masses (or Bonnor-Ebert masses; $M_{BE}
\propto M_J$), \citet{pado02,pado04} proposed that the CMF is set by
dividing the total available gas mass at each density into unstable
cores. \citet{pado07} propose that the peak mass in the CMF is given
by $M_c = 3 M_{BE,0}/M_A^{1.1}$ for $M_A \equiv v_{\mathrm{turb}}/v_A$
the Alfv$\acute{\mathrm{e}}$n Mach number in a cloud, and $M_{BE,0}$
the Bonnor-Ebert mass evaluated at the mean cloud density $n_0$. Here,
$v_A \equiv B/(4 \pi \rho)^{1/2}$ is the Alfv$\acute{\mathrm{e}}$n
speed.  For realistic mean GMC density $n_0 \sim 100 \pcc$ and
$\mathcal{M}_A \sim 1 -4$, from Equation (\ref{m_be}) the Padoan et
al formula in fact yields $M_c > 15 \Msun$; only if one chooses a much
higher reference density does this agree with observations. For the
unmagnetized case, \citet{pado07} propose that $M_c = 4
M_{BE,0}/\mathcal{M}^{1.7}$.
% where $\mathcal{M} \equiv v_{\mathrm{turb}} / c_s$ is the sonic Mach number. 
\citet{henn08} point out that shock compression is underestimated in the
magnetized case by \citet{pado07}, and advocate a formula similar to their
unmagnetized one: $M_c \sim M_{BE,0}/\mathcal{M}^{3/2}$. Since $\mathcal{M}
\gtrsim 10$ in massive GMCs, these formulae yield more realistic values
$M_c \sim\Msun$. Neither the \citet{pado07} or the \citet{henn08} proposal
has, however, been tested directly using self-gravitating numerical 
simulations.

In this contribution, we present results on core formation and
evolution based on a large suite of 3-dimensional numerical
simulations. Each simulation models a localized region of a turbulent
cloud in which there is an overall convergence in the velocity
field. Under the assumption that there is a dominant convergence
direction locally, we choose inflow along a single axis, so that
convergence is planar. With the more realistic geometry afforded by
the current simulations, we are able to check the results obtained by
\citet{gong09} for core building and collapse in supersonic flows. We
are also able to explore how the characteristic core mass is related
to the velocity of the converging flows. Since the speed of converging
flow is assumed to reflect the amplitude of the largest-scale
(dominant) motions in a GMC, this relates the characteristic core mass
to the turbulent Mach number in its parent GMC.  Although a number of
previous studies of core formation have been conducted, the present
investigation is distinguished by our systematic study of Mach number
dependence, together with our focus on internal structure and
kinematics of the cores that form.

The plan of this paper is as follows: In Section 2 we provide a
physical discussion of self-gravitating core formation in the
post-shock dense layers, identifying the mass, size, and time scales
expected to be important. In Section 3, we summarize the governing
equations and methods used in our numerical simulations. Section 4
describes the development of core structure and evolution in our
models, paying particular attention to the influence of Mach number
$\mathcal{M}$ on the evolution, and comparing collapse of individual
cores with \citet{gong09}. Section 5 describes our method of
core-finding, in which the largest closed contour of the gravitational
potential determines the core size. We demonstrate that this method
can be used for both three dimensional and two dimensional data with
similar results, and can thus be applied to find cores in observed
clouds. Section 6 describes the relations between core properties
(core mass, core radius and core collapse time) and the large-scale
Mach number of the converging flow, relating to the expectations from
gravitational instability discussed in Section 2. In Section 6, we
also quantify core shapes, and explore the relationship between core
structure and kinematics.  Section 7 summarizes our new results and
discusses our findings in the context of previous theories and
observations.

\section{The characteristic core mass and size}
Prior to describing our numerical model prescription and results, it
is useful to summarize the scales that are likely to be relevant for
formation of self-gravitating cores in GMCs. We shall assume
approximately isothermal conditions, consistent with observations
\citep[e.g.][]{blit07}.  The isothermal sound speed at a temperature
$T$ is
\begin{equation}\label{isosp_def}
c_s = 0.20 \kms \left(\frac{T}{10\K}\right)^{1/2}.
\end{equation}
If the density within clouds were uniform, the spatial scale relevant
for gravitational instability would be the Jeans length
\begin{equation}\label{LJ_def}
L_J \equiv c_s \left(\frac{\pi}{G \rho_{0}}\right)^{1/2} =2.76 \pc
\left(\frac{n_{H,0}}{10^2\pcc}\right)^{-1/2}
\left(\frac{T}{10\K}\right)^{1/2},
\end{equation}
evaluated at the mean density $\rho_0$. The corresponding Jeans mass is
\begin{equation}\label{MJ_def}
M_J \equiv \rho_0 L_J^3 = c_s^3 \left(\frac{\pi^3}{G^3\rho_0}\right)^{1/2}
=72
\Msun \left(\frac{n_{H,0}}{10^2 \pcc }   \right)^{-1/2}  
\left(\frac{T}{10\K  }  \right)^{3/2}.
\end{equation}
Note that $\rho_0(L_J/2)^3$ or $\rho_0 4 \pi (L_J/2)^3/3$ is sometimes
used for the Jeans mass.  The Bonnor-Ebert mass (eq. \ref{m_be}) for
$P_\mathrm{edge}=P_0\equiv\rho_0 c_s^2$ is $M_\mathrm{BE}=0.22
M_J(\rho_0)$.  The Jeans time at the mean cloud density is
\begin{equation}\label{tJ_def}
t_J \equiv \frac{L_J}{c_s} = \left(\frac{\pi}{G \rho_0}\right)^{1/2}=3.27\, t_{\mathrm{ff}}(\rho_0)=
1.4\times 10^7 \yr\, \left(\frac{n_{H,0}}{10^2\pcc}\right)^{-1/2}.
\end{equation}
We shall use the Jeans length, mass, and time at the unperturbed density as
our code units of length, mass, and time: $L_0 = L_J$, 
$M_0 = M_J$, and $t_0=t_J$.

Of course, GMCs are highly inhomogeneous, with core formation taking
place in the overdense regions that have the shortest gravitational
times. If the overdense regions within GMCs are produced by shocks in
the turbulent, supersonic flow, their density, and therefore the mass
scale and length scale for growth of self-gravitating structures, will
be related to the shock strength. Strongly magnetized shocks have less
compression than weakly magnetized shocks (while both will be present
in a turbulent flow), so we concentrate on the latter case.

If gravitationally unstable cores develop only in gas that has been
strongly compressed by shocks, the actual bounding pressure will be
much larger than $P_0 = \rho_0 c_s^2$.  In particular, an isothermal
shock with Mach number $\mathcal{M}$ will produce a post-shock region
with pressure $P_\mathrm{post-shock} = \rho_0 v^2 = \mathcal{M}^2
\rho_0 c_s^2 \gg P_0$. Thus, if cores preferentially form in
stagnation regions between shocks of Mach number $\mathcal{M}$, then
one can define an effective Bonnor-Ebert mass for these core-forming
regions within the turbulent flow by setting $P_\mathrm{edge} =
P_\mathrm{post-shock}$ in equation (\ref{m_be}):
\begin{equation}\label{m_be_ps}
M_{BE,\mathrm{post-shock}} \equiv 1.2\frac{c_s^3}{(G^3 \rho_0)^{1/2}} \frac{1}{\mathcal{M}} 
= 2.8\Msun \left( \frac{v}{1 \mathrm{km s^{-1}}}\right)^{-1} 
\left(\frac{n_{H,0}}{10^2 \pcc} \right)^{-1/2} \left(\frac{T}{10\K}\right)^{2}.
\end{equation}
The above simple argument suggests $M \propto v^{-1} \rho_0^{-1/2}T^2$ for the 
minimum mass of a star that forms via collapse of a core in a turbulent cloud
with velocity dispersion $v$, mean density $\rho_0$, and temperature $T$. 

Equation (\ref{m_be_ps}) provides a mass scale for fragmentation
within post-shock regions, but in fact instabilities take some time to
develop. Thus, it is useful to consider the evolution of a simple
system consisting of a planar shocked layer formed by a converging
flow \citep[see e.g.][]{elme78,lubo93,vish94,whit94,iwas08}.

For inflow Mach number $\mathcal{M}$, the surface density of the
post-shock layer at time $t$ is
\begin{equation}\label{surfd_t}
\Sigma(t) = \rho_0\,(v_{z,+}-v_{z,-})\,t = 2\rho_0\,\mathcal{M}\,c_s\,t,
\end{equation}
where $v_{z,+}$ and $v_{z,-}$ are the upward and downward converging
velocities.  If the sheet is not vertically self-gravitating, its
half-thickness is $H = \Sigma(t)/2\,\rho_p$ where $\rho_p \approx
\rho_0 \mathcal{M}^2$ is the post-shock density. The
non-self-gravitating half-thickness is thus
\begin{equation}\label{nsg_h}
H_\mathrm{nsg} \approx \frac{2 \rho_0\, \mathcal{M}\,c_s\,t}{2\rho_0\,\mathcal{M}^2} = \frac{c_s\,t}{\mathcal{M}}.
\end{equation}
As the surface density of the sheet increases, self-gravity will
become increasingly important in confining the gas. In the limit of
hydrostatic equilibrium, the height approaches
\begin{equation}\label{sg_h}
H_\mathrm{sg} = \frac{c_s^2}{\pi G\Sigma} = 
\frac{c_s}{2 \pi G \rho_0 \mathcal{M}t }.
\end{equation}
Note that the transition from non-self-gravitating ($H_{\rm
  nsg}\propto t$) to self-gravitating ($H_{\rm sg} \propto t^{-1}$)
  occurs at a time near
\begin{equation}\label{t_sg}
t_\mathrm{sg} \equiv \frac{1}{\left( 2 \pi G \rho_0 \right) ^{1/2}} = 0.22 t_J,
\end{equation}
defined by the condition $H_\mathrm{sg} = H_\mathrm{nsg}$.

The dispersion relation for in-plane modes in a slab, allowing for
non-zero $H$ \citep[e.g.][]{kim02}, is
\begin{equation}\label{disp_rl}
\omega^2 \approx c_s^2 k^2 - \frac{2\pi G \Sigma k}{1 + kH}.
\end{equation}
For the critical mode $\omega^2 =0$, so that 
\begin{equation}\label{disp_r_crit}
k_\mathrm{crit}H\,(1+k_\mathrm{crit}H) = 2 \pi H\, 
\frac{G\Sigma}{c_s^2} = 2 \pi \frac{H}{L_{\mathrm{J,2D}}},
\end{equation}
where 
\begin{equation}\label{LJ_2d}
L_{\mathrm{J,2D}} \equiv \frac{c_s^2}{G\Sigma}
\end{equation}
is the Jeans length for an infinitesimally-thin layer.  The solution
to equation (\ref{disp_r_crit}) is
\begin{equation}\label{k_crit_nsg_h}
k_{\mathrm{crit}}=\frac{2\pi}{L_{\mathrm{J,2D}}}
\frac{2}{1 + \left(1+8\pi \frac{H}{L_{\mathrm{J,2D}}} \right)^{1/2}}
=\frac{4 \pi G \rho_0 t \mathcal{M}}{c_s}
\frac{2}{1 + \left(1+8\pi \frac{H}{L_{\mathrm{J,2D}}} \right)^{1/2}},
\end{equation}
so that
\begin{equation}\label{lambda_crit}
\lambda_{\mathrm{crit}} = L_{\mathrm{J,2D}}
\frac{1 + \left(1+8\pi \frac{H}{L_{\mathrm{J,2D}}} \right)^{1/2}}{2}
=\frac{c_s}{2 G \rho_0 t \mathcal{M}}
\frac{1 + \left(1+8\pi \frac{H}{L_{\mathrm{J,2D}}} \right)^{1/2}}{2}. 
\end{equation}
The corresponding critical mass $(\lambda_\mathrm{crit}/2)^2 \Sigma$ is
\begin{equation}\label{crit_m}
M_{\mathrm{crit}} 
\equiv \frac{c_s^3}{32G^2 \rho_0 \mathcal{M} } 
\frac{\left[1+(1+8 \pi \frac{H}{L_\mathrm{J,2D}})^{1/2}\right]^2}{t}.
\end{equation}
Note that $H/L_\mathrm{J,2D}$ initially increases in time, during the
non-self-gravitating stage ($H_\mathrm{nsg}/L_\mathrm{J,2D} = 2 G
\rho_0 t^2$), and then approaches a constant
($H_\mathrm{sg}/L_\mathrm{J,2D} = 1/ \pi$).  At any time, all 
wavelengths $\lambda > \lambda_{\rm crit}$ 
have $\omega^2<0$, so that overdense regions of the corresponding sizes and
masses $M> M_{\rm crit}$ grow relative to their surroundings.

During the non-self-gravitating stage, the critical mass has a minimum
value at time
\begin{equation}
t_{\mathrm{crit,nsg,min}} = \left(\frac{3}{16 \pi G \rho_0} \right)^{1/2} 
= 0.14 t_J = 0.61 t_\mathrm{sg}
\end{equation}
given by
\begin{align}\label{crit_m_min_nsg}
\lefteqn{M_{\mathrm{crit,nsg,min}}
  = \frac{3\sqrt{3\pi}}{8} \frac{c_s^3}{(G^3\rho_0)^{1/2}} \frac{1}{\mathcal{M}}} \\
&&  = 3.0 \Msun \left( \frac{v}{1 \mathrm{km s^{-1}}}\right)^{-1} 
    \left( \frac{n_{H,0}}{10^2 \cm^3} \right)^{-1/2}
    \left( \frac{T}{10 \K} \right)^2.
\end{align}
The numerical coefficient in equation (\ref{crit_m_min_nsg}) is 1.15;
note that this is almost the same as in equation (\ref{m_be_ps}).

At late time, the critical mass from Equation (\ref{crit_m}) with
$H_\mathrm{sg}/L_\mathrm{J,2D} = 1/\pi$ becomes
\begin{equation}\label{m_crit_sg0}
M_\mathrm{crit,sg} =\frac{c_s^3}{2 G^2 \rho_0 \mathcal{M} t } = \frac{c_s^4}{G^2 \Sigma}.
\end{equation}
Expressing $M_\mathrm{crit,sg}$ in terms of the virial parameter 
$\alpha_\mathrm{vir}=5 \sigma_v^2R/GM_\mathrm{GMC}$ of the GMC, and using $\sigma_v = \mathcal{M} c_s$
and $M_\mathrm{GMC} = \pi R^2 \Sigma_\mathrm{GMC} = 4\pi R^3 \rho_0/3$, we have
\begin{equation}\label{m_crit_sg1}
M_\mathrm{crit,sg} = \left(\frac{3\pi \alpha_\mathrm{vir}}{20} \right)^{1/2} 
\frac{c_s^3}{(G^3 \rho_0)^{1/2}} \frac{1}{\mathcal{M}} \frac{\Sigma_\mathrm{GMC}}{\Sigma}.
\end{equation}
Here $\sigma_v$ is the large-scale one-dimensional velocity dispersion
in GMCs, which will be responsible for the largest scale, strongest
shocks.  Taking $\alpha_\mathrm{vir}= 2$, the coefficient in Equation
(\ref{m_crit_sg1}) is $0.97$, so this is very similar to equations
(\ref{m_be_ps}) and (\ref{crit_m_min_nsg}) if $\Sigma \sim
\Sigma_\mathrm{GMC}$. In dimensional units, the critical mass (for
$\Sigma = \Sigma_\mathrm{GMC}$) is
\begin{equation}\label{m_crit_sg3}
M_\mathrm{crit,sg}=2.5\Msun \left( \frac{v}{1 \mathrm{km s^{-1}}}\right)^{-1} 
\left(\frac{n_{H,0}}{10^2 \pcc} \right)^{-1/2} \left(\frac{T}{10\K}\right)^{2}.
\end{equation}

As noted above, equations (\ref{m_be_ps}), (\ref{crit_m_min_nsg}) --
(20) and (\ref{m_crit_sg1}) -- (\ref{m_crit_sg3}) all have a similar
form. An important task for numerical simulations is therefore to test
the hypothesis that the characteristic mass scale of collapsing cores
formed in turbulent, self-gravitating GMCs follows this scaling, i.e.
\begin{equation}\label{m_c_mdf0}
M_c=\psi \frac{c_s^4}{(G^3 \sigma_v^2\rho_0)^{1/2}} = 
\psi \times 2.6 \Msun \left( \frac{\sigma_v}{1 \mathrm{km s^{-1}}} \right)^{-1}
\left(\frac{n_{H,0}}{10^2 \pcc} \right)^{-1/2} \left(\frac{T}{10\K} \right)^2,
\end{equation} 
where $\psi$ is a dimensionless coefficient.

The critical mass given above is the smallest mass that can collapse,
given infinite time.  Since the growth rate depends on scale (and is
formally zero for critical perturbations), at any finite time only
cores that have grown sufficiently rapidly will be nonlinear enough to
collapse. It is therefore useful to consider how much growth has
occurred at a given time. Consider a perturbation of wavenumber $k$
that instantaneously has $\mathrm{d}^2 \delta \Sigma /\mathrm{d} t^2 =
- \omega^2 \delta \Sigma$ so that $\delta \Sigma = \delta
\Sigma_\mathrm{init} e^{\Gamma}$ where $\Gamma = \mathrm{ln}\,(\delta
\Sigma/\delta \Sigma_\mathrm{init}) = \int (-\omega^2)^{1/2}\,
\mathrm{d}t$.  Using equation (\ref{disp_rl}),
\begin{equation}\label{gamma0}
\Gamma = \int_{t_\mathrm{min}}^t (-\omega^2)^{1/2} \mathrm{d} t = 
\int_{t_\mathrm{min}}^t \left(\frac{2\pi G \Sigma k}{1+k H} - c_s^2 k^2 \right)^{1/2} \mathrm{d} t,
\end{equation}
where $t_\mathrm{min}$ is the instant when $\Sigma$ is large enough
that perturbations of wavenumber $k$ start to grow ($-\omega^2
\geqslant 0$).  With $\Sigma = 2 \rho_0 c_s \mathcal{M} t$,
$t_{\mathrm{min}} = c_s k (1 + kH)/(4 \pi G \rho_0 \mathcal{M})$. If we
assume $kH \ll 1$ (see below), then
\begin{equation}\label{gamma2}
\Gamma = \frac{2\sqrt{2}}{3} \kappa^{1/2} (\tau - \kappa/2)^{3/2},
\end{equation}
where $\kappa = k c_s/\sqrt{2 \pi G \rho_0 \mathcal{M}}$ and 
$\tau = t \sqrt{2 \pi G \rho_0 \mathcal{M}}$.

At a given time $t$ (or $\tau$) during the evolution, the mode $k_m$
(or $\kappa_m$) that has grown the most has $\partial \mathrm{ln}
\Gamma /\partial k=0$, which gives
\begin{equation}\label{gamma3}
\kappa_m = \frac{\tau}{2},
\end{equation}
and $\Gamma_\mathrm{max} = \Gamma(k_m) = \sqrt{3} \kappa_m^2 = \sqrt{3} \pi G 
\rho_0 \mathcal{M} t^2/2$.
The mass of this most-amplified mode is:
\begin{equation}\label{m_mgrn0}
M_m \equiv \left(\frac{\lambda_m}{2}\right)^2 \Sigma =
\left(\frac{2 \sqrt{3}\pi}{\Gamma_\mathrm{max}}\right)^{1/2} \frac{c_s^3}{(G^3\rho_0)^{1/2}} 
\frac{1}{\mathcal{M}^{1/2}},
\end{equation} 
where the time is
\begin{equation}\label{t_mgrn}
t = \left(\frac{2 \Gamma_\mathrm{max}}{\sqrt{3} \pi}\right)^{1/2} 
    \left(\frac{1}{G \rho_0 }\right)^{1/2}\frac{1}{\mathcal{M}^{1/2}},
\end{equation}
and $k_m = (\Gamma_\mathrm{max}/\sqrt{3})^{1/2} (2\pi
G\rho_0\mathcal{M})^{1/2}/c_s$, so that
\begin{equation}\label{lambda_mgrn}
\lambda_m=\left(\frac{2\sqrt{3}\pi}{\Gamma_\mathrm{max}}\right)^{1/2}
\frac{c_s}{(G \rho_0)^{1/2}}\frac{1}{\mathcal{M}^{1/2}}.
\end{equation}
With $\Gamma_\mathrm{max}=1$, the numerical coefficient for $M_m$ in
Equation (\ref{m_mgrn0}) is $3.30$, and Equation (\ref{t_mgrn}) gives
$t = 0.34 t_J \mathcal{M}^{-1/2}$, corresponding to $\tau = 1.5$.
Note that for low Mach number, this time exceeds $t_\mathrm{sg}$ (see
eq. \ref{t_sg}), whereas for high Mach number it does not.  Also, note
that with $H < \cs t_{\rm sg}/{\cal M} \equiv H_{\rm max}$ (see
eqs. \ref{nsg_h} - \ref{t_sg}), $k_m H < k_m H_{\rm max} = \Gamma_{\rm
max}^{1/2}(\sqrt{3}\mathcal{M})^{-1/2}$.  Taking $\Gamma_{\rm max}=1$,
$k_m H < 0.8$ for $\mathcal{M} > 1$, with $k_m H \ll 1$ for  
$\mathcal{M} \gg 1$.  This verifies self-consistency of 
the assumption made in obtaining equation (\ref{gamma2}).

Written in terms of $v, \rho_0$, and $T$, the most-amplified mass is
\begin{equation}\label{m_mgrn2}
M_m= 19.1\Msun \left( \frac{v}{1 \mathrm{km s^{-1}}}\right)^{-1/2} 
\left(\frac{n_{H,0}}{10^2 \pcc} \right)^{-1/2} \left(\frac{T}{10\K}\right)^{7/4}
\left(\Gamma_\mathrm{max}\right)^{-1/2}.
\end{equation}
Comparing equation (\ref{m_mgrn2}) with equation (\ref{m_crit_sg3}),
we see that a different dependence on velocity (or Mach number) is
expected for the first core to collapse (equation \ref{m_mgrn2}), compared
to the typical core to form eventually (equation  \ref{m_crit_sg3}).
Similar results to equation (\ref{m_mgrn0}) have previously been
discussed by other authors. \citet{whit94} point out that the
fastest-growing scale $\sim L_\mathrm{J,2D} \sim c_s/(G \rho_0
\mathcal{M}  t)$ will become nonlinear if the time exceeds the growth time
$\sim L_\mathrm{J,2D}/c_s \sim (G\rho_0 \mathcal{M} t )^{-1}$, which
occurs for $t \sim (G\rho_0 \mathcal{M})^{-1/2}$ (cf. our
eq. \ref{t_mgrn}). This corresponds to a length scale
$L_\mathrm{fragment} \sim c_s (G \rho_0 \mathcal{M})^{-1/2}$ (cf. our
eq. \ref{lambda_mgrn}), and a mass scale $M_\mathrm{fragment} \sim
c_s^3 (G^3 \rho_0 \mathcal{M})^{-1/2}$ (cf. our eq. \ref{m_mgrn0}). By
direct integration of the perturbation equation of the converging-flow
system, \citet{iwas08} find that the fastest-growing mode becomes
nonlinear at time $0.96 \delta_0^{-0.1} (G \rho_0
\mathcal{M})^{-1/2}$, for $\delta_0$ the initial amplitude (cf. our
eq. \ref{t_mgrn}, which has a coefficient 0.6 if
$\Gamma_\mathrm{max}=1$).

Finally, we note that the characteristic mass scale at late times
given in equation (\ref{m_c_mdf0}) can be connected to observed core
mass scales using the empirical relationships among turbulence level,
size, and mass for GMCs.  In terms of the viral parameter
$\alpha_\mathrm{vir} \equiv 5\sigma_v^2 R/(GM_{\rm GMC})$ and the GMC
surface density $\Sigma_\mathrm{GMC} \equiv 4 \rho_0 R/3$, equation
(\ref{m_c_mdf0}) can be re-expressed as
\begin{equation}\label{m_c_mdf1}
M_c=1.5\psi \frac{c_s^4}{\alpha_\mathrm{vir}^{1/2}G^2\Sigma_\mathrm{GMC}}=
\psi\times 1\Msun \left(\frac{T}{10K}\right)^2 
\left(\frac{\Sigma_{\mathrm{GMC}}}{100\Msun \mathrm{\pc^{-2}}} \right)^{-1} 
\alpha_{\mathrm{vir}}^{-1/2}.
\end{equation}
With $\alpha_\mathrm{vir} \sim$ 1 -- 2 and $\Sigma_\mathrm{GMC} \sim
100 \Msun \pc^{-2}$ for observed clouds \citep{solo87,mcke07,heye09},
the mass scale is intriguingly similar to the characteristic (peak)
mass of CMFs within nearby molecular clouds.  This relation
potentially also offers a prediction for the peak of the CMF (and
ultimately the IMF) when stars form under conditions different from
those in most Milky Way GMCs. In particular, high temperature (up to
$\sim 70\K$) may hold in starburst regions where the radiation field
is strong and turbulent dissipation rates are high; since the
temperature dependence of equation (\ref{m_c_mdf1}) is steeper than
the dependence on surface density, this could imply higher masses
under those conditions.

\section{Methods for numerical simulations}
The numerical simulations we present here are conducted with the
\emph{Athena} code (\citealt{gard05,gard08,ston08, ston09}), using the
HLLC solver \citep{toro99} and second order reconstruction
\citep{ston08}.  To calculate the self-gravity of our slab domains,
which are periodic in-plane and open in the $z$ direction, the Fast
Fourier Transformation (FFT) method developed by \citet{koya09} is
used.  We solve the three-dimensional equations of hydrodynamics,
\begin{equation}\label{cont_eq}
\frac{\partial \rho}{\partial t} + \mathbf{\nabla}
     \cdot (\rho\,\textbf{\em v}) =0,
\end{equation}

\begin{equation}\label{mom_eq}
\frac{\partial \textbf{\em v}}{\partial t} + 
   \textbf{\em v} \cdot \mathbf{\nabla}\textbf{\em v} = -\frac{\mathbf{\nabla}P}{\rho} - 
   \mathbf{\nabla}\Phi;
\end{equation}
and the Poisson equation,
\begin{equation}\label{poisson_eq}
\mathbf{\nabla}^2 \Phi = 4 \pi G \rho,
\end{equation}
where $\Phi$ is the gravitational potential.  The isothermal
assumption $P = c_s^2 \rho$ is adopted. \citet{pavl06} found the
isothermal approximation is adequate for simulations of the
interstellar medium even with strong turbulence, which implies strong
shocks in GMCs.

The code unit of density $\rho_{0}$ is a fiducial density representing
the volume-averaged ambient density in a cloud on large scales; this
characterizes the mean density of converging flows. For the code unit
of velocity, we adopt the isothermal sound speed $c_s$ (see
eq. \ref{isosp_def}). For the unit of length, we adopt $L_0 = L_J$,
the Jeans length at the fiducial density (see eq. \ref{LJ_def}).  The
mass and time units for the simulation are then $M_0 = M_J$ (see
eq. \ref{MJ_def}) and $t_0=t_J$ (see eq. \ref{tJ_def}).

In making comparison to observations, the total surface density
integrated through the domain
\begin{equation}\label{surfd_def}
\Sigma = \int \rho(x,y,z) dz = \Sigma_0 \int \frac{\rho}{\rho_0} \frac{dz}{L_J}
\end{equation}
is useful, for $\Sigma_0 \equiv \rho_0 L_J = 9.49 \Msun \pc^{-2} 
(T/10K)^{1/2}(n_{H,0}/10^2 \pcc)^{1/2}$.
In terms of the column density of hydrogen,
\begin{equation}\label{col_den_h}
N_H = \frac{\Sigma}{1.4 m_p} = N_0 \int \frac{n_{H}}{n_{H,0}} \frac{dz}{L_J}
\end{equation}
for $N_0 \equiv n_0 L_J=8.51 \times 10^{20} \cm^{-2} 
(T/10K)^{1/2}(n_{H,0}/10^2 \pcc)^{1/2}$.
The mean line-of-sight velocity is calculated by
\begin{equation}\label{v_los}
\langle v_\mathrm{los}\rangle = \frac{\int \rho v_\mathrm{los}
ds}{\int \rho ds},
\end{equation}
and the corresponding dispersion of $\langle v_\mathrm{los}\rangle$ is
defined as
\begin{equation}\label{sigma_los}
\sigma_\mathrm{los}^2 = \frac{\int \rho(v_\mathrm{los} -\langle v_\mathrm{los}\rangle)^2 ds}{\int \rho ds},
\end{equation}
where $ds = sec\theta\, dz$ and $\theta$ is the tilt angle of the
observer with respect to the $z$ axis.

Our model prescription consists of a converging flow augmented with
turbulent velocity perturbations.  In our parameter survey, the Mach
number $\mathcal{M}$ of the inflow velocity ranges from 1.1 to
9. Thus, two flows converge toward the central plane $z=0$ from the
upper $z$-boundary (with mean velocity $-\mathcal{M}c_s$) and the
lower $z$-boundary (with mean velocity $\mathcal{M}c_s$). The initial
density is uniform and set to $\rho_0$, and the density at the
inflowing $z$-boundaries is also set to $\rho_0$ throughout the
simulation. The boundaries in the $x$ and $y$ directions are periodic.

For both the whole domain initially and the inflowing gas
subsequently, we apply perturbations following a Gaussian random
distribution, with a Fourier power spectrum of the form
\begin{equation}\label{pert_pow_eq}
  \langle \left| {\delta \textbf{\em v}_k} \right|^2 \rangle \propto k^{-2},
\end{equation}
for $|kL/2 \pi| < N/2$, where $N$ is the resolution and $L$ is the
size of the simulation box in $x$ and $y$. The power spectrum is
appropriate for supersonic turbulence as observed in GMCs
\citep{mcke07}.  The perturbation velocity fields are pre-generated
with resolution $256^3$ in a box of size $L^3$. The perturbation
fields are advected inward from the $z$-boundaries at inflow speed
$\mathcal{M}\,c_s$: at time intervals $\Delta t = \Delta
z/(\mathcal{M}c_s)$, slices of the pre-generated perturbation fields
for $v_x, v_y$ and $v_z$ are read in to update values in the ghost
zones at the $z$-boundaries.

In addition to exploring dependence on the mean inflow Mach number
$\mathcal{M}$, we also test dependence on the amplitude of turbulent
perturbations on top of this converging flow.  From the scaling law
\citep[see e.g.,][]{lars81,heye04} of self-gravitating molecular
clouds, $\delta v(l) \propto l^{1/2}$, we can write the velocity
dispersion at scale $l$ in terms of cloud-scale one-dimensional
velocity dispersion $\sigma_v$ and cloud radius $R$ as $\delta
v_{1D}(l) = \sigma_v (l/2R)^{1/2}$.  The velocity dispersion at the
scale of the simulation box $L$ is
\begin{equation}\label{sig_1d_box}
\frac{\delta v_{1D}(L)}{c_s} = 
\frac{\sigma_v}{c_s} \left( \frac{L}{2R} \right)^{1/2}
=\frac{\sigma_v}{c_s}\left(\frac{L}{L_J}\right)^{1/2} \left(\frac{2R}{L_J}\right)^{-1/2}.
\end{equation}
In terms of the viral parameter $\alpha_\mathrm{vir} \equiv
5\,\sigma_v^2 R/(GM)$, where $M = 4 \pi R^3 \rho_0/3$ is the cloud
mass, the ratio between $\sigma_v$ and $c_s$ is
\begin{equation}\label{sig_1d_cs_r}
\frac{\sigma_v}{c_s} = 2 \pi \left( \frac{\alpha_\mathrm{vir}}{15} \right)^{1/2} \frac{R}{L_J}.
\end{equation}
Solving equation (\ref{sig_1d_cs_r}) for $2R/L_J$ and substituting
into equation (\ref{sig_1d_box}), we have the amplitude of
perturbation for the simulation box:
\begin{equation}\label{pert_ampl}
\frac{\delta v_{1D}(L)}{c_s} = 
\left(\frac{\alpha_\mathrm{vir} \pi^2}{15}\right)^{1/4} \left(\frac{\sigma_v}{c_s}\right)^{1/2}
\left(\frac{L}{L_J}\right)^{1/2}.
\end{equation}
Thus, if the size of the simulation box is $L = L_J$ and $\alpha_{vir}
= $ 1 -- 2, the perturbation amplitude would be
\begin{equation}\label{pert_ampl_f}
 \frac{\delta v_{1D}(L_J)}{c_s} \approx \left(\frac{\sigma_v}{c_s}\right)^{1/2}.
\end{equation}
If we take the Mach number of the inflow, $\mathcal{M}$, as comparable
to the value $\sigma_v/c_s$ of the whole cloud, then equation
(\ref{pert_ampl_f}) implies that higher converging velocities would be
associated with higher amplitudes for the perturbation fields, for a
given simulation box size $L_J$.  To test the influence of the
perturbation amplitude, we conduct two sets of simulations with $10\%$
and $100\%$ of the value $\delta
v_{1D}(L_J)=(\mathcal{M}/3)^{1/2}c_s$. Hereafter, we denote these
cases as low amplitude and high amplitude initial perturbations,
respectively.

For each Mach number $\mathcal{M}$ at each amplitude, we run 20
simulations with different random realizations of the same
perturbation power spectrum, in order to collect sufficient
statistical information on the core properties that result. The whole
set of simulations therefore consists of 180 separate runs. The
resolution for low amplitude perturbation simulations is $N_x \times
N_y \times N_z = 256\times256\times 96$, with domain size $L_x \times
L_y \times L_z/L_J^3 = 1\times1\times0.375$; for high amplitude the
resolution is $N_x \times N_y \times N_z = 256\times256\times 160$,
with domain size $L_x \times L_y \times L_z/L_J^3 =
1\times1\times0.625$.  The domain in the $z$ direction is smaller than
in the $x$ and $y$ directions since the reversed shock generated by
the inflow only propagates a relatively short distance and the
post-shock dense layer is thin, i.e., the basic geometry remains
planar. The domain in the $z$ direction is large enough so that 
the post-shock layer does not evolve to reach the $z$
boundaries.

We note that our assumption of perturbed velocities but uniform
densities in the inflowing gas is not fully realistic, since the flow
entering a strong shock within a GMC will in general have internal
density structure.\footnote{ Other recent simulations of post-shock
structure formation in converging flows have similarly assumed uniform
density for the inflow (see e.g.
\citealt{heit08,2009MNRAS.398.1082B}, and references therein).}  In
fact, the velocity perturbations we introduce do lead to moderate
(order-unity) density fluctuations, as we have found by conducting
comparison simulations with self-gravity turned off.  These density
fluctations are what seed the growth of self-gravitating structures.
The main emphasis of the current work is to investigate how the
development of self-gravitating structures depends on the inflow Mach
number, which sets the mean density (and hence the gravitational
timescale) in the post-shock layer; previous studies have not tested
the Mach number dependence of gravitational fragmentation.  By varying
the velocity perturbation amplitudes of the inflow, we have begun to
explore the effect of pre-existing density structure on
self-gravitating core development in shocked regions.  This
exploration can be extended and made more realistic (in terms of
upstream structure) by investigating internal evolution of shocked
layers within larger fully-turbulent clouds having a range of mean
Mach number; we are currently pursuing a numerical study along these
lines.  The models presented here may be thought of as investigating
self-gravitating structure growth within the first strong shocks to
develop inside a cloud.

\section{Development of structure and core evolution} 
As discussed in Section 1, \citet{gong09} proposed a unified model for
core formation and evolution in supersonic turbulent
environments. Based on spherical-symmetry numerical simulations, four
stages were identified: core building, core collapse, envelope infall
and late accretion.  The duration of each stage, and the structure and
kinematics of cores at varying stages were also analyzed. While the
comparison of those results to observations is very encouraging, the
assumption of spherical symmetry is clearly unrealistic. One of the
key goals of this work is to check if core building and collapse still
develop in a similar manner when the spherical-symmetry assumption is
relaxed.  Because the time step becomes very short in late stages, we
halt the simulations; thus the current models do not address envelope
infall and late accretion stages.

\begin{figure}[ht]
\centerline{
\includegraphics[width=0.8\textwidth,angle=90]{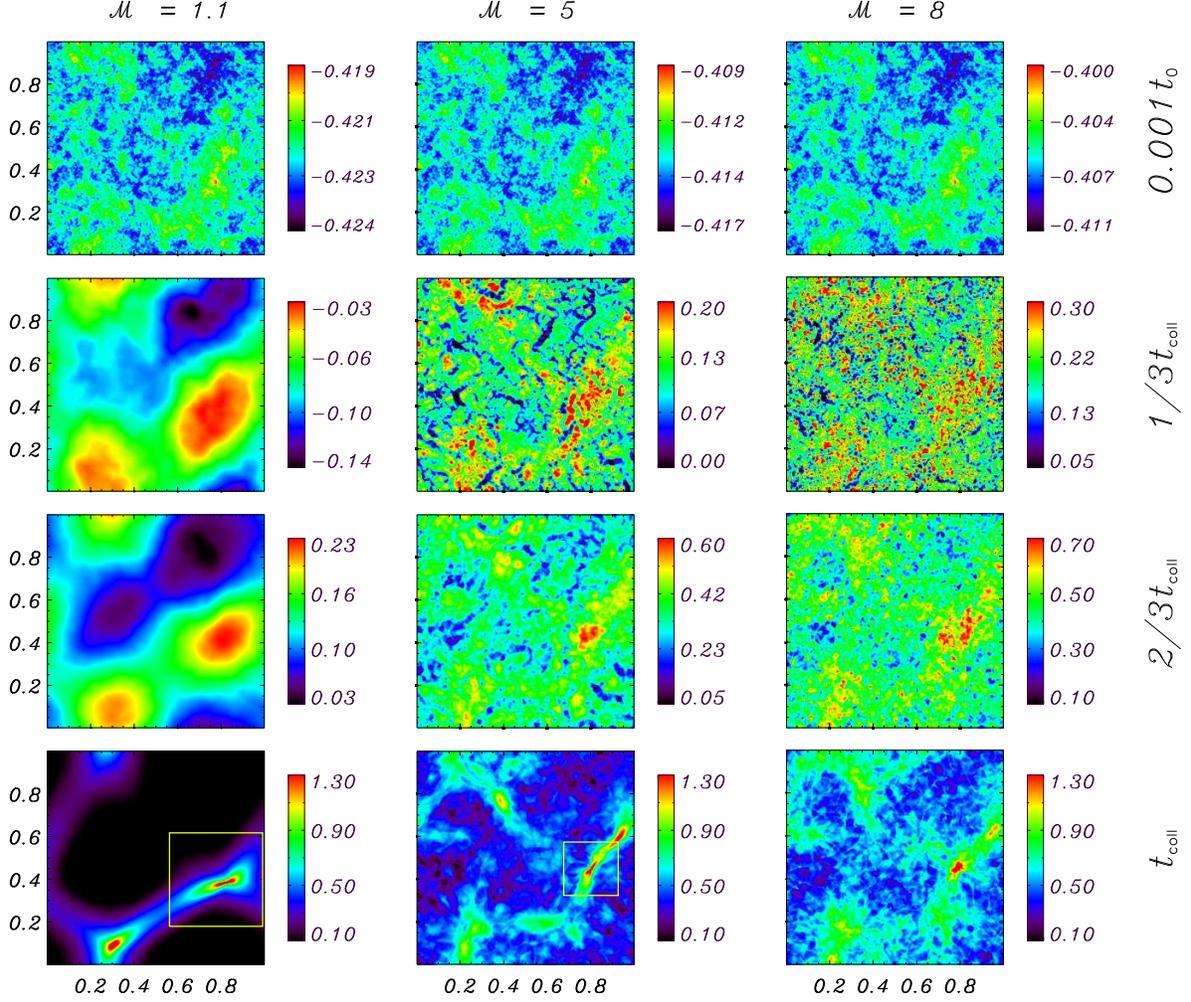}
}
\caption{Evolution of surface density projected in the $z$ direction
(color scale $\mathrm{log} \Sigma/\Sigma_0 = \mathrm{log} N/N_0$; see
eqs. \ref{surfd_def}, \ref{col_den_h}) for converging-flow Mach
number $\mathcal{M} = 1.1$ (left column), $\mathcal{M} = 5$ (middle
column) and $\mathcal{M} = 8$ (right column) models with the same
initial perturbation patterns. The four panels from top to bottom in
the each column show surface density snapshots at four instants: $t =
0.001\,t_0$, $1/3 t_{\mathrm{coll}}$, $2/3 t_{\mathrm{coll}}$, and
$t_{\mathrm{coll}}$, with $t_{\mathrm{coll}}$ the duration of the
whole simulation. These three simulations have $10\%$ initial
perturbation amplitude (see eq. \ref{pert_ampl}). The values of
$t_{\mathrm{coll}}$ are $0.636 t_0, 0.280 t_0$ and $0.232 t_0$ for
$\mathcal{M}=1.1,5$ and $8$ respectively (see eq. \ref{tJ_def} for
definition of $t_0$).  Cores are clearly smaller and more irregular
for high-$\mathcal{M}$ models. The squares indicate the most evolved
cores for $\mathcal{M} = 1.1$ and 5.  }
\label{fig:comp_mach}
\end{figure}

Figure \ref{fig:comp_mach} shows evolution of the surface density
(eq. \ref{surfd_def}) for models with $\mathcal{M} = 1.1$ (left
column), $\mathcal{M} = 5$ (middle column) and $\mathcal{M} = 8$
(right column), all with same realization for the perturbation
velocities.  The top panel of each column shows the surface density
very early on; the patterns are identical but the amplitudes are
different. The bottom panel shows the surface density when the most
evolved core collapses for each case.  Hereafter we shall use
$t_{\mathrm{coll}}$ to denote the total time to reach collapse of the
most evolved core, in terms of the code unit $t_0$ (eq. \ref{tJ_def}).
The four images from top to bottom in the same column show the surface
density at four instants: $t = 0.001\,t_0$, $1/3 t_{\mathrm{coll}}$,
$2/3 t_{\mathrm{coll}}$, and $t_{\mathrm{coll}}$. Note that
$t_{\mathrm{coll}} = 0.636 t_0, 0.280 t_0$ and $0.232 t_0$ for the
$\mathcal{M}=1.1,5$ and $8$, respectively. These three simulations
have low initial perturbation amplitude (cf. eq. \ref{pert_ampl_f}).

From Figure \ref{fig:comp_mach}, two features are immediately
apparent. First, the input perturbation field patterns determine the
later structural evolution and there is a ``family resemblance'' for
the models at different Mach number. This is because the post-shock
dense layer retains a memory of the perturbation velocity fields in
the direction parallel to the plane of the layer since $v_x$ and $v_y$
are unchanged across the shock interface. Comparing the first plot to
the last plot of each column, cores form in regions where the density
perturbation amplitudes are initially higher than the surroundings as
a result of convergence in the $x-y$ plane. These overdense regions
develop into long, thin filaments, within which cores grow and then
collapse.

Second, the specific properties of cores, such as the total number and
individual volumes (as well as their masses), are determined by
$\mathcal{M}$. The dense cores for $\mathcal{M} = 1.1$ are smoother
than the cores for $\mathcal{M} = 8$, and they cover larger
areas. During the middle and late stages of evolution, more small
scale filamentary structures are evident in the higher Mach number
cases. At a given scale, the input $v_x$ and $v_y$ perturbations are
higher for larger $\mathcal{M}$, with the resulting compressions
making more prominent ``burrs'' around cores. The ``burrs'' are also
less smoothed for the high Mach number cases, because the shorter
free-fall time at the higher post-shock density means that the core
collapses sooner. Thus, as the velocity of the converging flow and
additional perturbations increases, the result is smaller, denser,
more irregular, and more ``hairy'' cores.

\begin{figure}[ht]
\centerline{
\includegraphics[width=0.9\textwidth,angle=0]{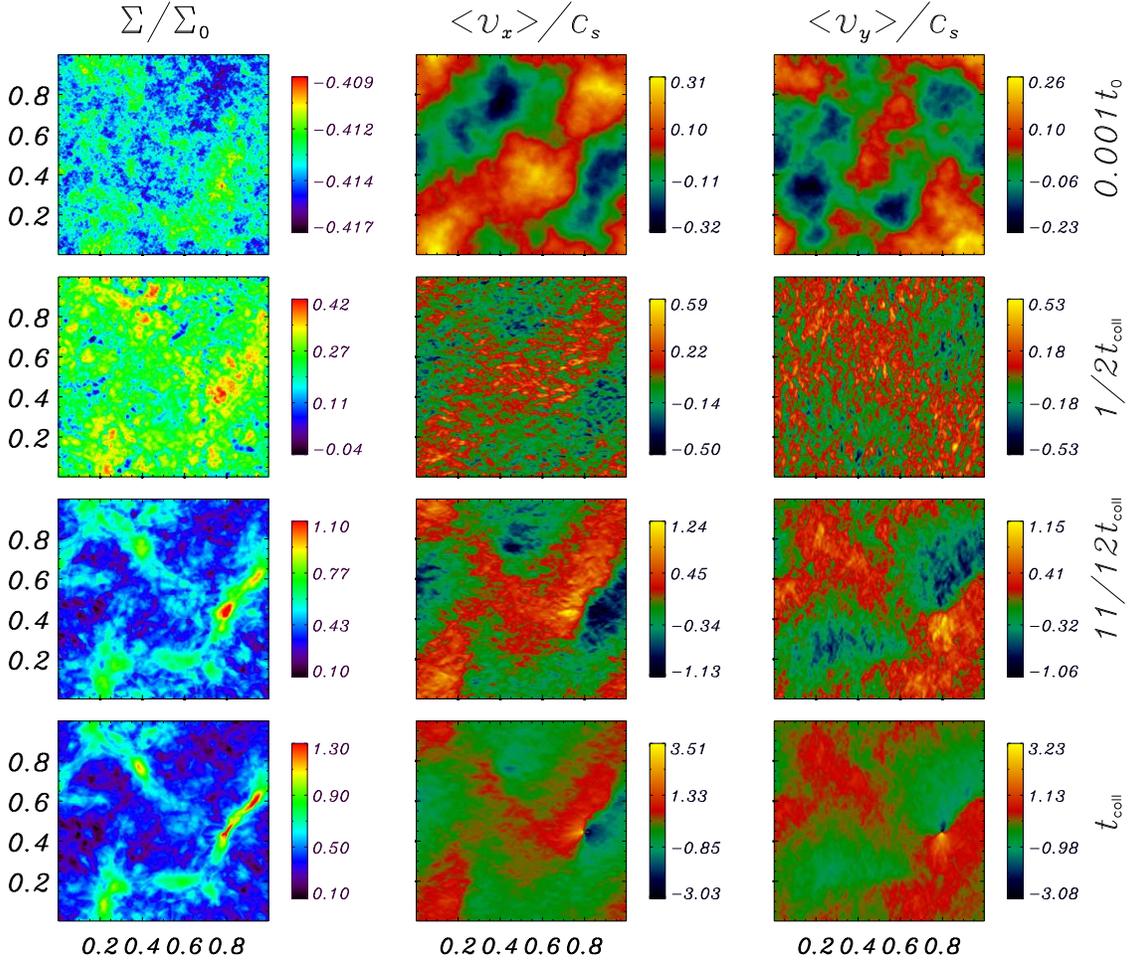}
}
\caption{Evolution of surface density (left column, log color scale)
and the in-plane velocity components $\left<v_x\right>$ (middle
column) and $\left<v_y\right>$ (right column) projected in the $z$
direction for the Mach number $\mathcal{M} = 5$ model shown in
Figure \ref{fig:comp_mach}, where $\left<v\right>=\int \rho v dz/\int
\rho dz$.  The four panels from top to bottom in the each column show
four instants: $t = 0.001\,t_0$, $1/2 t_{\mathrm{coll}}$, $11/12
t_{\mathrm{coll}}$, and $t_{\mathrm{coll}}$, with $t_{\mathrm{coll}} =
0.28 t_0$ the duration of the simulation (see eq. \ref{tJ_def} for
definition of $t_0$). In-plane velocity fields are initially low, but
grow to become supersonic, creating filaments that fragment into
cores.  }
\label{fig:vx_vy_proj}
\end{figure}

Figure \ref{fig:vx_vy_proj} shows evolution of surface density and the
mean in-plane velocities $\left<v_x\right>$ and $\left<v_y\right>$ for
the $\mathcal{M} =5$ model shown in Figure \ref{fig:comp_mach}. The
mean velocities are calculated by $\left< v\right> = \int \rho v dz
/\int \rho dz$ with $v=v_x$ or $v_y$. The left column shows surface
density, and the middle and the right columns show $\left<v_x\right>$,
$\left<v_y\right>$ respectively. At early stages, only scattered high
surface density spots appear. The large-scale spatial correlation of
these overdense regions is evident, however, even at early times. The
mean velocities also have small amplitudes at early stages. The
large-scale converging (in-plane) velocity regions that eventually lead to the
most prominent filaments are already evident from the first frames,
however.  At late stages, the overdense regions start to collect into
filaments. The converging (in-plane) velocities grow due to self-gravity of the
forming filaments; in addition, purely hydrodynamic instabilities
(such as the nonlinear thin-shell instability, e.g.
\citealt{vish94,2007ApJ...665..445H}) in the shock-bounded layer may
enhance early growth of perturbations.\footnote{ We have conducted
comparison tests of selected models without self-gravity, finding that
surface density fluctuations can grow to order-unity level in high
Mach number cases.}  When converging in-plane flows become supersonic,
discontinuities in the density and velocity develop. These sharp
fronts, as well as the collapsing motions centered on the most evolved
cores, are evident in Fig. \ref{fig:vx_vy_proj} at $t=11/12\,
t_\mathrm{coll}, t_\mathrm{coll}$.

Thus, we see that turbulent motions even at sub-pc scales seed the
growth of structures, and self-gravity reinforces and amplifies these
motions. The growth of dense cores and larger scale filaments is
simultaneous, both a consequence of turbulence and self-gravity.

Similar to our results in \citet{gong09} for spherical symmetry, we
find that core building lasts most of the time up to
$t_{\mathrm{coll}}$, while the core collapse itself is rapid for the
most evolved cores.  Defining the ``supercritical'' period as the
stage at which $\rho_{\mathrm{center}} /\rho_{\mathrm{edge}} \geq 10$
for the most evolved core, this first occurs at $0.589\,t_0,
0.240\,t_0$ and $0.209\,t_0$ respectively for the $\mathcal{M} = 1.1,
5$ and $8$ models shown in Figure \ref{fig:comp_mach} (we note that
$\rho_{\mathrm{edge}}$ is close to the post-shock density). Taking the
difference with $t_\mathrm{coll}$, $\Delta t_{\mathrm{supcrit}}/t_0 =
0.047, 0.040$ and $0.023$. From \citet{gong09}, the supercritical
stage lasts about $10\%$ of $t_{\mathrm{coll}}$ for cores found in
shocked converging spherical flows.  For the three cases shown here,
$\Delta t_{\mathrm{supcrit}}/t_{\mathrm{coll}}$ is $7 \%, \, 14\%,$
and $10\%$, consistent with our previous results.  The core building
stage lasts about $90\%$ of $t_{\mathrm{coll}}$.

To express $\Delta t_{\mathrm{supcrit}}$ in terms of observables, we
renormalize using the mean core density $\rho_{\mathrm{mean}}$ at the
instant of collapse. This quantity, $\Delta
t_{\mathrm{supcrit}}/t_{ff}(\rho_{\mathrm{mean}})=\Delta
t_{\mathrm{supcrit}}/t_0\times 3.27
(\rho_{\mathrm{mean}}/\rho_0)^{1/2}$ is measured to be $0.9,2.1$ and
$0.8$ for $\mathcal{M} = 1.1, 5$ and $8$ respectively; i.e. $\Delta
t_\mathrm{supcrit}$ is comparable to $t_{ff}(\rho_\mathrm{mean})$.
The values of $\Delta t_{\mathrm{supcrit}}$ are $6.6\times10^5 \yr,
5.6\times10^5 \yr$ and $3.2\times10^5 \yr$ for $\mathcal{M} = 1.1, 5$
and $8$ respectively, if we take the inflowing ambient medium density
as $n_{H,0} = 100 \cm^{-3}$; these are reduced to $2\times10^5 \yr,
1.7\times10^5 \yr$ and $1\times10^5 \yr$ for $n_{H,0} = 1000
\cm^{-3}$.

\begin{figure}[ht]
\centerline{
\includegraphics[width=0.6\textwidth,angle=90]{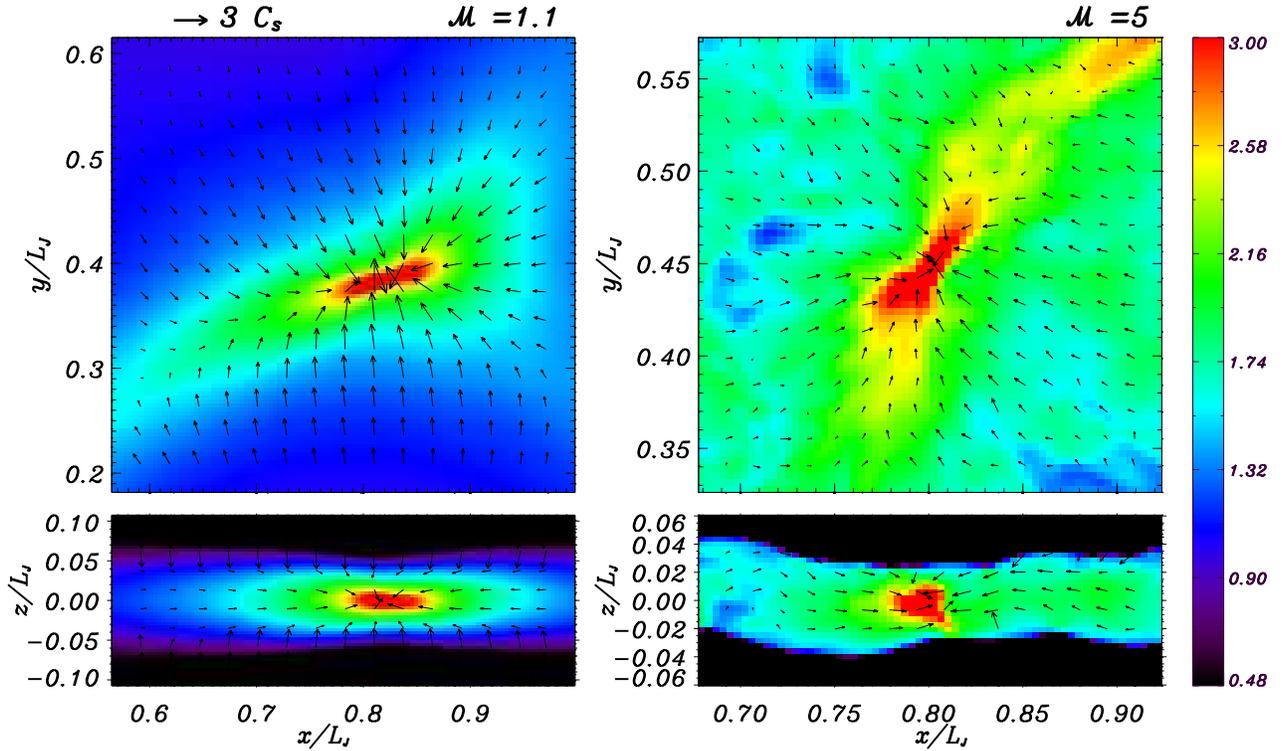}
}
\caption{Density and velocity field cross-sections at the time
$t_\mathrm{coll}$ in the most evolved core, for $\mathcal{M} = 1.1$
(left column) and $\mathcal{M}=5$ (right column). These correspond to
the most evolved cores (as indicated with boxes) in Figure
\ref{fig:comp_mach} for $\mathcal{M} = 1.1, 5$ respectively. The color
scale represents $x-y$ and $x-z$ slices through the volume density
($\mathrm{log}\rho/\rho_0$).  The direction and length of arrows
indicate the direction and magnitude of the local velocity, with scale
as indicated in the upper left. At this stage of collapse, velocities
increase toward the center.  }
\label{fig:core_structure_2d}
\end{figure}

Figure \ref{fig:core_structure_2d} shows the cross-sections of the
density and velocity field across the center of the most evolved cores
(the locations of these cores are indicated in Figure
\ref{fig:comp_mach}) for $\mathcal{M} = 1.1, 5$ during the late
collapse phase.  The instants of the plot for $\mathcal{M} = 1.1, 5$
are $0.625\,t_0$ and $0.273\,t_0$ respectively.  The top panels show
the $x-y$ cross-section of density and velocity vectors composed of
$v_x$ and $v_y$ in the same plane. The bottom part shows the $x-z$
cross-section and velocity vectors composed of $v_x$ and $v_z$. The
velocity field clearly shows inward collapse. The amplitudes of the
velocity field are smaller in the outer part and larger in the inner
part, indicating the core is at a very late stage of the
``outside-in'' collapse.

\begin{figure}[ht]
\centerline{
\includegraphics[width=0.7\textwidth,angle=90]{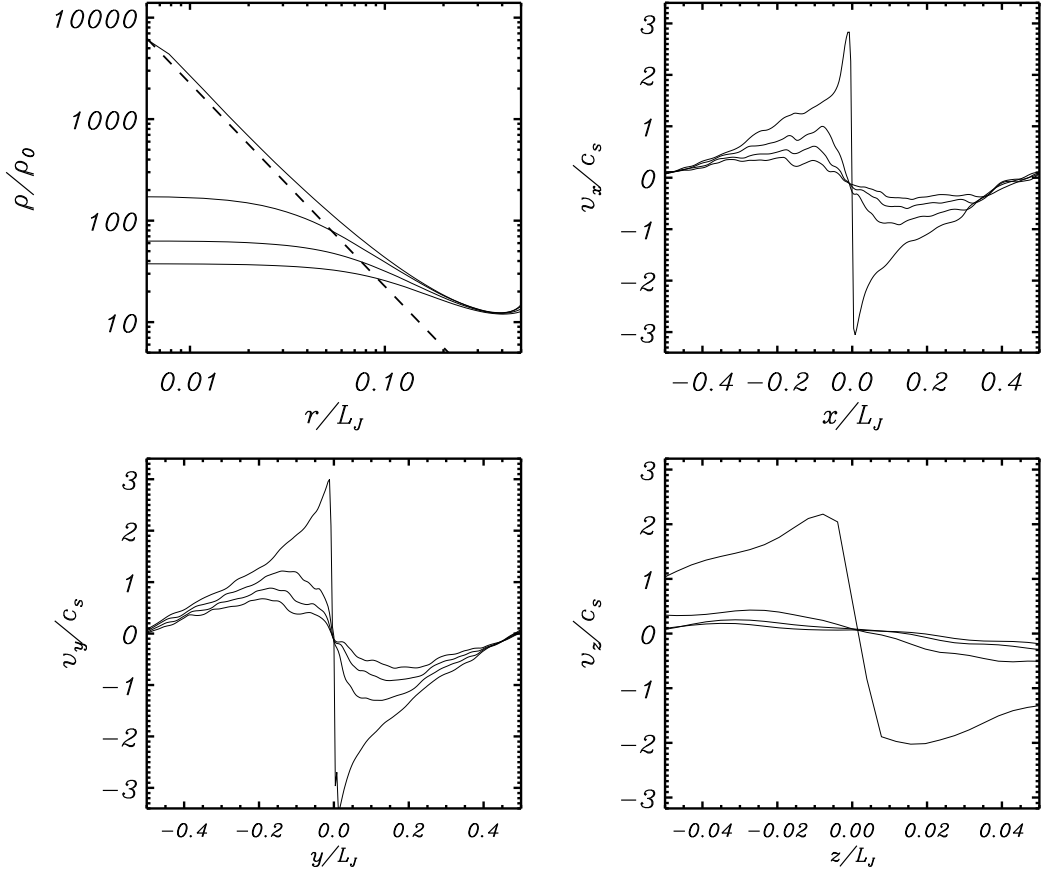}
}
\caption{Radial density and velocity profiles during collapse, for the
most evolved core shown in Figure \ref{fig:comp_mach} and Figure
\ref{fig:core_structure_2d} for $\mathcal{M} = 1.1$.  The density
profiles are averaged azimuthally in the $x-y$ plane about the center
of the core.  The dashed line is the Larson-Penston asymptotic density
profile $\rho/\rho_0=8.86(r/L_J)^{-2}/(2\pi)^2$ (i.e. 
$\rho = 8.86 c_s^2/[4 \pi G r^2]$).  The other three
plots show the corresponding velocity profiles versus distance in
the $x, y$ and $z$ direction, respectively. The instants shown are
$0.549\,t_0, 0.576\,t_0, 0.603\,t_0, 0.632\,t_0 \approx
t_\mathrm{coll}$, with the most evolved profiles in each case having
the largest excursions.  The collapse develops in an ``outside-in''
manner with the maximum in $v$ moving inward with time.  The density
profile approaches the Larson-Penston profile with time.  }

\label{fig:core_structure_1d0}
\end{figure}

\begin{figure}[ht]
\centerline{
\includegraphics[width=0.7\textwidth,angle=90]{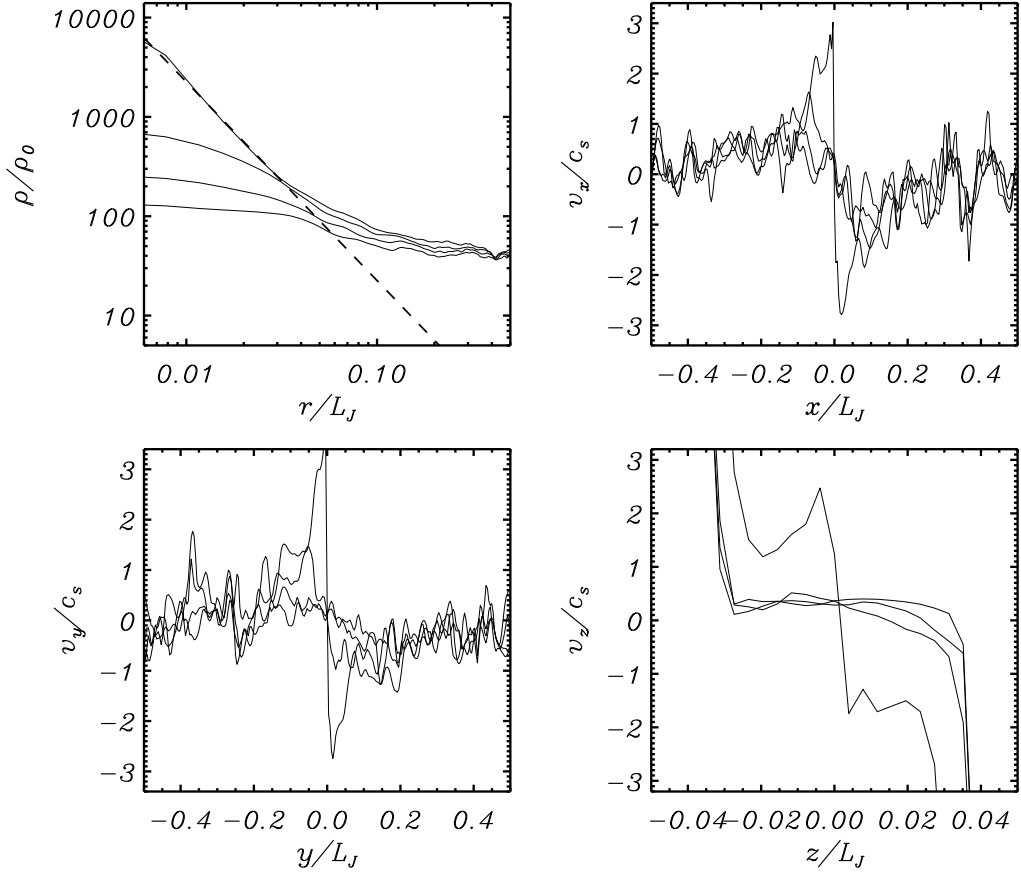}
}
\caption{Same as in Figure \ref{fig:core_structure_1d0} for the
most-evolved core of the $\mathcal{M}=5$ model shown in
Fig. \ref{fig:comp_mach}. The profiles are shown at $t =
0.219\,t_0,0.238\,t_0,0.257\, t_0,0.276\,t_0$, with the density at the
final time reaching the Larson-Penston solution.}
\label{fig:core_structure_1d1}
\end{figure}

Figure \ref{fig:core_structure_1d0} and Figure
\ref{fig:core_structure_1d1} show the evolution of the density and
velocity profiles of the cores in Figure \ref{fig:core_structure_2d}.
The density profiles are azimuthally-averaged over the $x-y$ plane.
The velocity profiles are along each cardinal axis
($\hat{x},\hat{y},\hat{z}$) through the core center. The instants for
the four profiles have equal intervals $0.027\,t_0$ for $\mathcal{M} =
1.1$ and equal intervals $0.019\,t_0$ for $\mathcal{M} =5$
respectively.  The first instant for both cases is subcritical
(i.e. $\rho_{\mathrm{center}} /\rho_{\mathrm{edge}} \leq 10$) and the
second instant is close to $t_{\mathrm{supcrit}}$. The dramatic
increase of the central density during collapse is clearly evident for
both cases, and the collapse develops in an ``outside-in'' manner with
the maximum in $v$ moving inward in time. The density profile
approaches the asymptotic ``Larson-Penston'' profile $\rho/\rho_0 =
8.86 (r/L_J)^{-2}/(2\pi)^2$ at the instant of central singularity
formation, and the in-plane velocities $v_x, v_y$ approach
$-3.3\,c_s$, which is the ``Larson-Penston'' limit.  Before the time
$t_{\mathrm{supcrit}}$ is reached, the velocity is subsonic throughout
the core region. For all of the simulations we have conducted, the
peak of the velocity profile becomes supersonic only at the very end
of the collapse stage, similar to the results shown here.

Overall, we conclude that the evolution of individual cores in these
3D simulations follows a similar progression to the
spherically-symmetric 1D simulations of \citet{gong09}. The core
building stage lasts over $90\%$ of the time to collapse, and cores
become more stratified over time.  The onset of the collapse is in an
``outside-in'' manner, and leads to a dramatic increase in the central
density. As a central singularity is approached, the density and
velocity profiles approach the ``Larson-Penston'' asymptotic
solution. These cores form and collapse within larger-scale filaments
that also grow in contrast over time.

\section{Core-finding method}
The algorithm adopted for core-finding can either subtly or more
seriously affect the core properties that result
\citep[e.g.][]{pine09}. The most commonly-used methods in
observational work are based on contouring column density or emission
intensity \citep[e.g. the popular \emph{Clumpfind} method
of][]{will94}. For theoretical work, density-contouring methods,
sometimes incorporating further tests to determine if a structure is
gravitationally bound, have frequently been used
\citep[e.g.][]{gamm03}. Here we shall instead use the gravitational
potential isosurfaces to identify cores. In very recent work,
\citet{smit09} took a similar approach, noting that one advantage of
the gravitational potential is that it yields smoother core boundaries
than the density. Another advantage is that the gravitational
potential connects more directly to the fundamental physics that
determines core evolution. During formation stages, self-gravity
gathers material to build up cores, and later it drives the collapse
of supercritical cores.

To identify cores via the gravitational potential, we first find and
mark all the local minima of the gravitational potential; second, we
find the largest closed potential contour (or isosurface) surrounding
each individual minimum. In the second step, we increase the contour
level from the bottom of a given potential well step by step until it
violates another minimum's marked territory. We define the region
enclosed by the largest closed contour as a core. The contour interval
$\Delta \Phi$ has negligible effect on the results as long as it is
small enough (typically $\leq 0.03 c_s^2$).  If the distance between
two minima is smaller than 10 pixels (corresponding to a physical
distance $\sim 0.03-0.1 \mathrm{pc}$ for $n_{H,0} \sim 10^2 - 10^3
\mathrm{cm}^{-3}$), the regions associated with these two minima are
merged and treated as a single core. Since we do not continue the
simulation after the most evolved core collapses, we apply the
algorithm to the last output from each simulation.

Since gas with sufficient thermal and kinetic energy need not be
permanently (or even temporarily) bound to a given core, the
gravitational potential is not the final word. The lower density outer
parts of a core are the most subject to loss. We can test this effect
on core identification by adding thermal energy to the gravitational
energy, and only assigning a given fluid element to a core if $E_{\rm
th} + E_g < 0$. For any fluid element, the specific thermal energy is
taken to be $E_{\rm th} = 3/2 c_s^2$, and the specific gravitational
potential energy is taken to be $E_g = \Phi - \Phi_{\rm max}$, where
$\Phi_{\rm max}$ is the potential of the largest closed contour that
defines the core. \footnote{We note that $|E_g|$ for a core embedded
within a dense filament (or sheet) may be much lower than $|E_g|$ for
the same core in isolation. In assessing whether a core is bound, it
is crucial to take tidal gravity effects into account. If these tidal
effects are neglected, $|E_g|$ will be overestimated by a factor $\sim
\Sigma_{\rm core}/(\Sigma_{\rm core} - \Sigma_{\rm filament})$, which
is quite large if the contrast between a core and its surroundings is
modest.}  Including a thermal energy condition in core definition
decreases the volume (or area in 2D) of the cores. Of course, the
thermal energy can in fact be radiated away, so that gas that is
initially near the largest closed contour may become more strongly
bound after the interior of a core collapses.  In this case, the
potential alone could determine the final core mass. Short of
following cores through the final stages of star formation, we
consider it useful to compare cores with and without a
thermal - gravitational energy criterion. 
Hereafter, we term our core-finding method
``gravitational identification'' (GRID). We refer to the region within
the largest closed gravitational potential isosurface surrounding each
local minimum as a GRID-core. For each GRID-core, the region which has 
$E_{\rm th} + E_g < 0$ is referred to as a bound GRID-core.

Because volume density data cubes are not directly accessible in
observations, three-dimensional gravitational potential contouring is
only applicable to model data from numerical simulations.  It is
therefore interesting to explore gravitational potential contouring of
surface density maps, which are direct observables. To identify cores
in a surface density map, we have to calculate the gravitational
potential first. For a layer of half-thickness $H$, the gravitational
potential component $\Phi_{\textbf{\em k},\,\mathrm{2D}}$ of surface
density component $\Sigma_{\textbf{\em k}}$ (Fourier transform of equation
(\ref{surfd_def})) in phase space is
\begin{equation}\label{psol}
\Phi_{\textbf{\em k},\,\mathrm{2D}} = -\frac{2 \pi G \Sigma_{\textbf{\em k}}}
{\left| {\textbf{\em k}}\right|(1 + \left|{\textbf{\em k}H}\right|)},
\end{equation}
where $\left|\textbf{\em k}\right| = \sqrt{k_x^2+k_y^2}$.  Note that
for $|\textbf{\em{k}}H| \gg 1$, $\Phi_{\textbf{\em k},\,\mathrm{2D}}
\sim -4 \pi G \rho_\textbf{\em{k}} /k^2$, which is the solution of the
Poisson equation in three dimensions, for $\rho_\textbf{\em k} =
\Sigma_\textbf{\em k}/2H$. For $|\textbf{\em{k}}H| \ll 1$,
eq. (\ref{psol}) is the solution of the Poisson equation for an
infinitesimally thin layer.  The gravitational potential
$\Phi_\mathrm{2D}(x,y)$ is the inverse Fourier transform of $\Phi_{\textbf{\em
k},\,\mathrm{2D}}$.  Given the 2D gravitational potential field
$\Phi_\mathrm{2D}(x,y)$, we can apply the GRID procedure as for 3D. In
Section 6, we will compare the results from GRID using $\Phi(x,y,z)$
and $\Phi_\mathrm{2D}(x,y)$ (using $H = \delta z$). Hereafter we use
``2D'' to denote the results from applying the GRID method to
surface density and ``3D'' for applying the GRID method to the volume density.

\begin{figure}[ht]
\centerline{
\includegraphics[width=0.7\textwidth,angle=90]{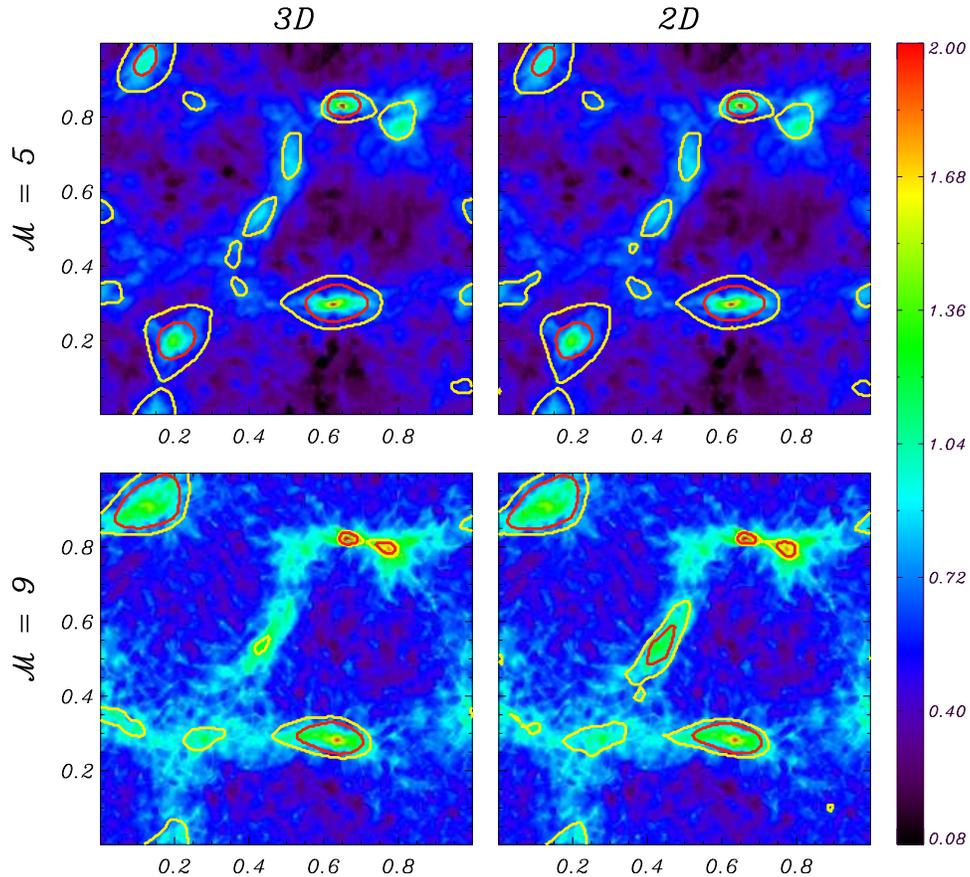}
}
\caption{Comparison of GRID-cores defined via the gravitational
potential computed from 3D volume density ($\Phi$, left column) and 2D
projected surface density ($\Phi_\mathrm{2D}$, right column). The top
row shows $\mathcal{M} = 5$ and bottom row $\mathcal{M} = 9$. The
areas enclosed by yellow curves are the GRID-cores determined by the
largest closed gravitational potential ($\Phi$ or $\Phi_\mathrm{2D}$)
contour surrounding a local potential minimum, and the areas enclosed
by red curves are the bound GRID-cores.  Color scale shows projected
surface density ($\mathrm{log} \Sigma/\Sigma_0$ ) in all panels.  
Cores identified using $\Phi$ and $\Phi_\mathrm{2D}$ agree quite well.  }
\label{fig:corefind_comp_2d3d}
\end{figure}

As an example, Figure \ref{fig:corefind_comp_2d3d} shows the
comparison of GRID-cores and bound GRID-cores between 3D and 2D for
$\mathcal{M} = 5$ and $9$. The top portion shows core areas identified
for the $\mathcal{M} =5$ model using $\Phi$ (top left) and
$\Phi_\mathrm{2D}$ (top right). The bottom portion shows the same
comparison for $\mathcal{M} = 9$ with cores found from $\Phi$ (bottom
left) and from $\Phi_\mathrm{2D}$ (bottom right). (Note that the
$\mathcal{M}=5$ and $\mathcal{M}=9$ simulations have the same initial
velocity perturbations patterns, which is why the overall structure is
similar). In all plots, the areas enclosed by yellow contours are the
GRID-cores and the areas enclosed by red contours are the bound
GRID-cores. The core areas for the 3D plots are the projection of the
3D core volume onto the $z=0$ plane.  For the $\mathcal{M}=5$ model,
the 2D and 3D core-finding procedures identify 12 and 13 cores
respectively; the cores and the bound regions are located at nearly
the same positions. For the $\mathcal{M}=9$ model, 7 cores are
identified for both cases. One bound core in 2D lacks a 3D
counterpart, implying the corresponding potential well in 3D is
too shallow (see discussion of potential well depths in Section 6).

\begin{figure}[ht]
\centerline{
\includegraphics[width=0.7\textwidth]{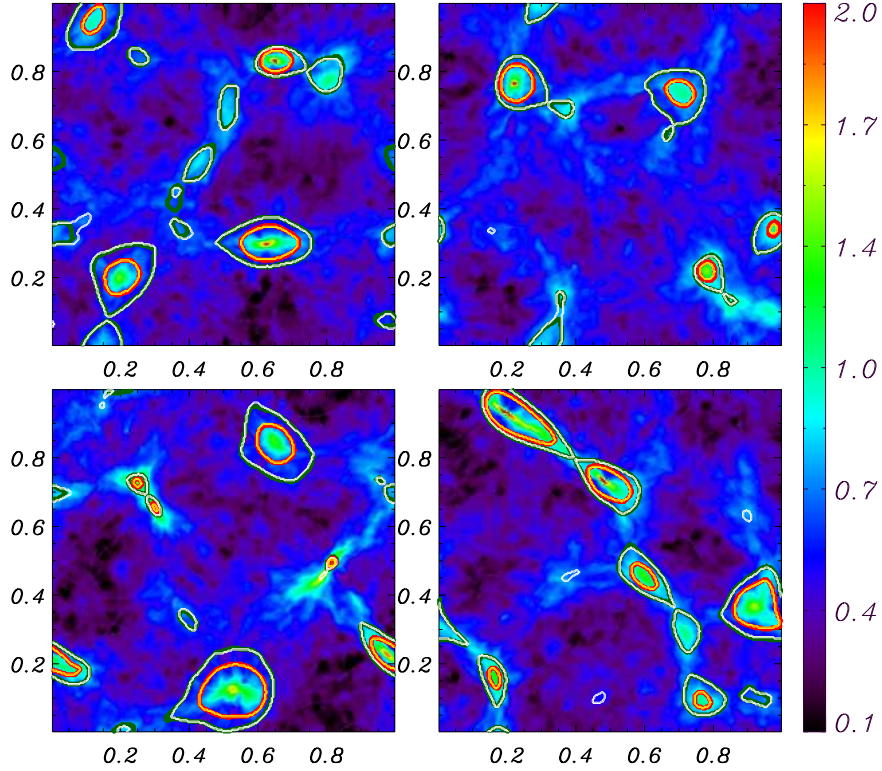}
}
\caption{Late stage surface density ($\mathrm{log} \Sigma/\Sigma_0$)
and GRID-core comparison for four different random perturbation
realizations of the $\mathcal{M}$ = 5 model. The snapshots are at
$t=0.282 t_0, 0.304 t_0, 0.304 t_0, 0.302 t_0$ from left to right and
top to bottom. The corresponding maximum densities are $1.0\times10^5
\rho_0, 1.53\times10^5 \rho_0, 8.18\times10^4 \rho_0, 1.34\times10^5
\rho_0$.  The white and green curves are GRID-cores defined by the
largest closed contour of the gravitational potential ($\Phi$ and
$\Phi_\mathrm{2D}$ respectively) surrounding each potential minimum.
The red and yellow curves are the bound GRID-cores obtained using
$\Phi$ and $\Phi_\mathrm{2D}$, respectively. Except for a few small,
shallow cores, the core-finding algorithms in 2D and 3D give quite
similar results.  }
\label{fig:comp_seed}
\end{figure}

In addition to finding almost all of the same core centers (defined by
the potential minimum), the areas marked by the 3D and 2D GRID algorithms
are almost the same. Figure \ref{fig:comp_seed} show the results of
GRID for four simulations for $\mathcal{M}=5$. The white contours mark
GRID-cores from 3D density and the green contours mark GRID-cores from
2D surface density. The red and yellow contours mark the bound
GRID-cores for 3D and 2D respectively. The areas identified for the
cores agree quite well.  Over all, we conclude that the 2D GRID
algorithm can give nearly identical core-finding areas as the 3D GRID
algorithm.

In spite of the overall similarity between 2D and 3D GRID-core
finding, there are minor differences in the results. In the each panel
of Figure \ref{fig:comp_seed}, a few GRID-cores in relatively low
density regions are identified in 2D but not in 3D. In
comparing core properties between 2D and 3D, we shall apply additional
resolution criteria to eliminate these small, shallow cores.

\section{Core properties} 

To obtain a sufficient statistical sample, we conduct 20 simulations
for each value of the Mach number ($\mathcal{M} =
1.1,2,3,4,5,6,7,8,9$) and compute GRID-core masses and radii for each
model (180 models total). Each of the 20 simulations for a given
$\mathcal{M}$ is perturbed by a different realization of the velocity
field. As an example of the differences with different random
realizations of the power spectrum, Figure \ref{fig:comp_seed} shows
the snapshots of surface density at a late stage for four
different $\mathcal{M}=5$ simulations. The 3D GRID core numbers are 9,
6, 9 and 7. The corresponding core mass ranges are [0.00151,\,0.158]
$M_0$, [0.0051,\,0.128] $M_0$, [0.0013,\,0.242] $M_0$ and
[0.031,\,0.250] $M_0$. The core numbers and core masses from
simulations with different seeds are in a similar range; the same is
true for cases with other Mach numbers.

The GRID-core masses for 3D and 2D are $M_{\mathrm{3D}} = \int \rho\
dxdydz$ and $M_{\mathrm{2D}} = \int \Sigma\ dxdy$, respectively. The
GRID-core radius for 3D is defined as the equivalent radius of a 3D
sphere with the same volume $V_\mathrm{3D}$: $r\,_{\mathrm{3D}} \equiv
(3V_\mathrm{3D}/4\pi)^{1/3}$.  The effective 2D GRID-core radius is
calculated from the area $S_\mathrm{2D}$ of the core region as:
$r\,_{\mathrm{2D}} \equiv (S_\mathrm{2D}/\pi)^{1/2}$. To ensure that
identified GRID-cores are numerically well-resolved, we only retain
cores with effective radii $\ge 4$ zones. We define a background
surface density as the mean of the bottom $10\%$ of the surface
density; this mean value can be subtracted from the surface density in
the core region when calculating $M_\mathrm{2D}$. As mentioned in
Section 2, a more restrictive definition includes only gas
with thermal plus gravitational energy negative; these bound GRID-cores 
are first identified by the gravitational potential, and then
pixels are excluded if the sum of thermal energy and gravitational
potential is greater than 0.

\begin{figure}[ht]
\centerline{
\includegraphics[width=0.7\textwidth,angle=90]{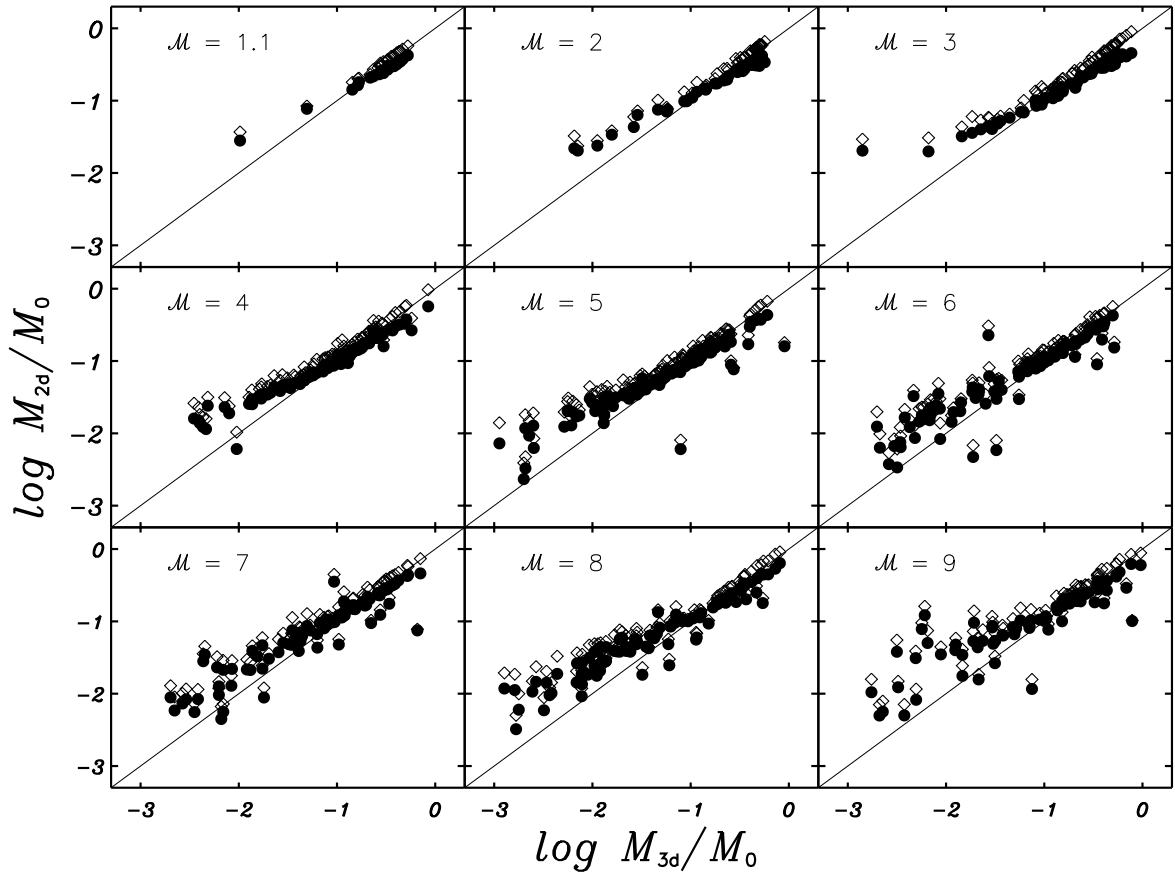}
}
\caption{GRID-core mass obtained from from 2D ($M_{\mathrm{2D}}$)
versus 3D ($M_{\mathrm{3D}}$).  Diamonds are $M_{\mathrm{2D}}$ for 2D
GRID-cores without background subtraction, and dots are
$M_{\mathrm{2D,bs}}$ for 2D GRID-cores with background subtraction.
The mass unit $M_0$ is given in equation (\ref{MJ_def}). Solid lines
represent $M_\mathrm{2D} =M_\mathrm{3D}$; higher-mass cores are
consistent with this.  }
\label{fig:2d_vs_3d_gb}
\end{figure}

Figure \ref{fig:2d_vs_3d_gb} shows $M_{\mathrm{2D}}$ versus
$M_{\mathrm{3D}}$ for GRID-cores, for each Mach number of the low
amplitude perturbation set. Note that only cores with same center of
the local potential minima are shown here.  Both 2D GRID-core masses
without background subtraction ($M_{\mathrm{2D}}$, diamonds in the
figure) and 2D GRID-core masses with background subtraction
($M_\mathrm{2D,bs}$, dots in the figure) are shown versus
$M_{\mathrm{3D}}$. For large masses, $M_{\mathrm{2D}}$ agrees well
with $M_{\mathrm{3D}}$ while $M_\mathrm{2D,bs}$ is slightly lower than
$M_\mathrm{3D}$. For small masses, $M_{\mathrm{2D,bs}}$ agrees better
than $M_\mathrm{2D}$ with $M_{\mathrm{3D}}$. Both $M_{\mathrm{2D}}$
and $M_\mathrm{2D,bs}$ agree with $M_{\mathrm{3D}}$ better for high
mass than low mass.

\begin{figure}[ht]
\centerline{
\includegraphics[width=0.7\textwidth,angle=90]{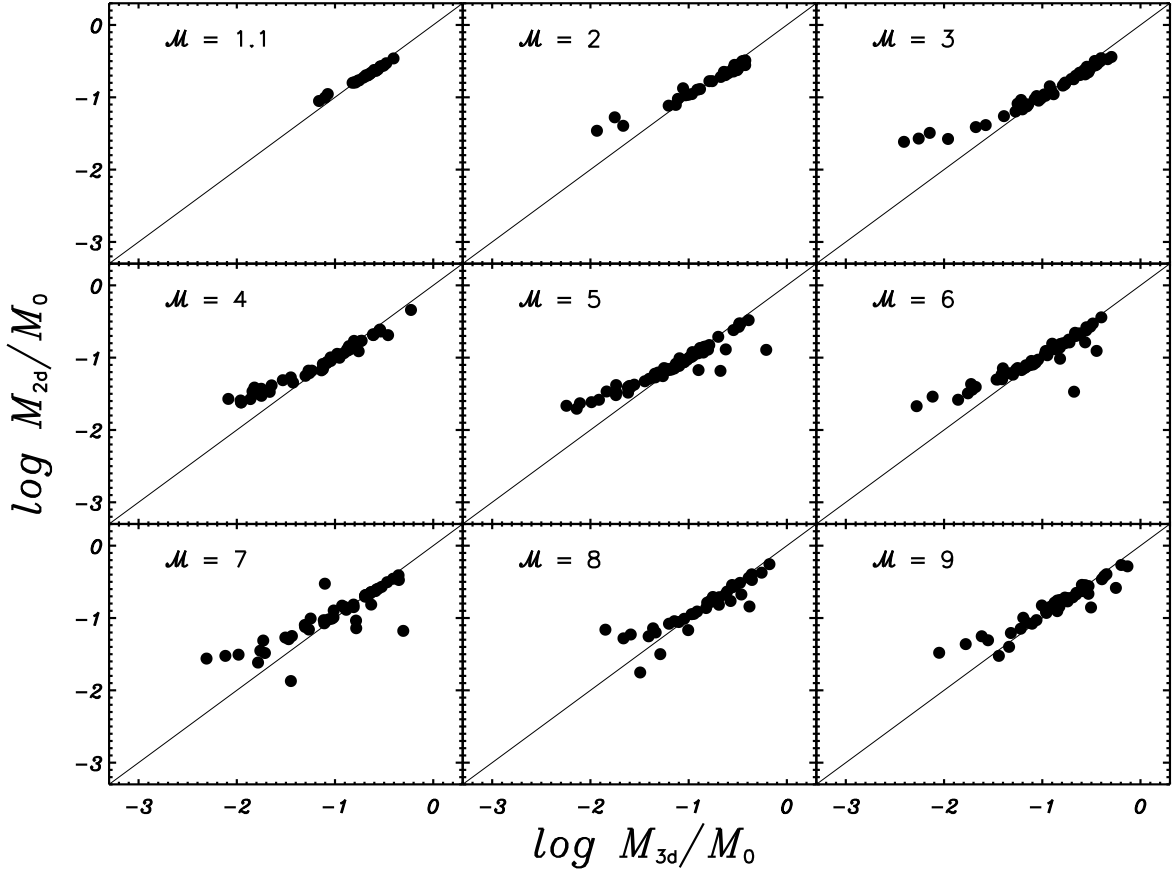}
}
\caption{Bound GRID-core mass for 2D with background subtraction
($M_{\mathrm{2D,bs,th}}$), versus bound GRID-core mass for 3D
($M_{\mathrm{3D,th}}$). When the condition $E_{\rm th} + E_g < 0$ is
included in the core definition, the lowest mass cores are eliminated
and $M_{\rm 2D,bs,th}$ agrees well with $M_{\rm 3D,th}$ down to $\sim
10^{-2}M_0$.  }
\label{fig:2d_vs_3d_gtbt}
\end{figure}

Figure \ref{fig:2d_vs_3d_gtbt} shows a similar comparison of bound
GRID-cores for 2D and 3D. The background surface density is subtracted
for 2D GRID-core masses, so that we show $M_{\mathrm{2D,bs,th}}$
versus $M_{\mathrm{3D,th}}$. Here, the subscript ``th'' represents
inclusion of a thermal energy criterion in defining bound GRID-cores,
which eliminates most of the small cores. At high masses,
$M_{\mathrm{2D,bs,th}}$ agrees with $M_{\mathrm{3D,th}}$ for bound
GRID-cores better than $M_\mathrm{2D,bs}$ agrees with
$M_{\mathrm{3D}}$ for the whole set of GRID-cores. This is because
only zones sufficiently near the potential minimum where 
$E_{\rm th} + E_g < 0$ are included in
bound GRID-cores; these regions are not sensitive to projection
effects. At low masses, $M_{\mathrm{2D,bs,th}}$ exceeds
$M_{\mathrm{3D,th}}$ for bound GRID-cores, meaning that imposing the
thermal - gravitational energy criterion affects $M_{\mathrm{3D,th}}$
more than $M_{\mathrm{2D,bs,th}}$.

To understand the difference between the 2D and 3D GRID-core masses,
we consider the shape of the gravitational potential well for surface
density and volume density. From equation (\ref{psol}),
$\Phi_{\mathrm{2D},k} \propto - k^{-1}$ whereas $\Phi_{\mathrm{3D},k}
\propto - k^{-2}$.  At larger $k$, corresponding to smaller scales,
$|\Phi_{3D}|$ decreases faster than $|\Phi_\mathrm{2D}|$. That means
that the small 2D GRID-cores cover more area than small 3D GRID-cores,
evident at the low end of each panel in
Fig. \ref{fig:2d_vs_3d_gb}. Also, gravitational potential wells of
middle-sized 2D GRID-cores are deeper than those of 3D middle-sized
GRID-cores. If the shallow parts of the potential are excluded by
applying a thermal energy requirement, 3D GRID-cores are affected more
than 2D GRID-cores.  Moderate-mass GRID-cores that have
$M_\mathrm{2D,bs}$ and $M_{\mathrm{3D}}$ comparable will thus have
$M_{\mathrm{3D,th}}$ lower than $M_{\mathrm{2D,bs,th}}$, as is evident
in Fig. \ref{fig:2d_vs_3d_gtbt}.  As mentioned in Section 5, we
include the term $|\textbf{\em{k}}|H$ to allow for the non-zero
thickness of the layer perpendicular to the plane. This can, in
principle, help decrease the gap between the 2D and 3D gravitational
potentials.  In practice, however, we find that the value for $H$ to
make the central-to-edge value of $\Phi_\mathrm{2D}$ comparable to
that for $\Phi$ is smaller than $\delta z$. Although the 2D and 3D
gravitational potentials are not exactly the same, Figure
\ref{fig:2d_vs_3d_gtbt} shows that 2D and 3D bound GRID-cores masses
are generally close down to $\sim 10^{-2} M_0$ (which is $\lesssim 1
\Msun$ for typical conditions, from eq. \ref{MJ_def}).

\begin{figure}[ht]
\centerline{
\includegraphics[width=0.7\textwidth,angle=90]{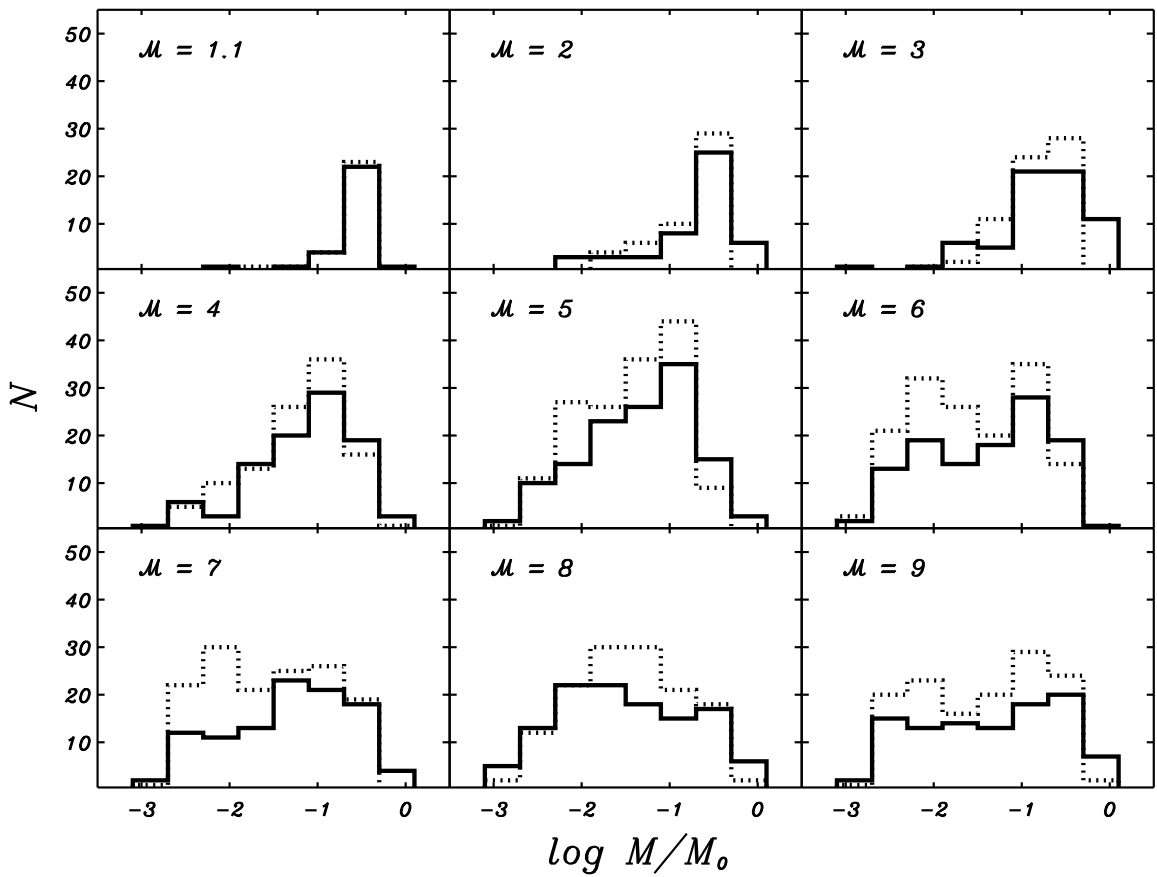}
}
\caption{Histograms of all GRID-core masses found in all simulations
for each Mach number $\mathcal{M}$ for low amplitude
perturbations. Solid lines are for 3D GRID-cores ($M_\mathrm{3D}$) and
dashed lines are for 2D GRID-cores with background subtraction
($M_\mathrm{2D,bs}$).
%Only the gravitational potential is used to define the cores. 
The 2D and 3D distributions are similar for all Mach numbers.}
\label{fig:hist_comp_l}
\end{figure}

\begin{figure}[ht]
\centerline{
\includegraphics[width=0.7\textwidth,angle=90]{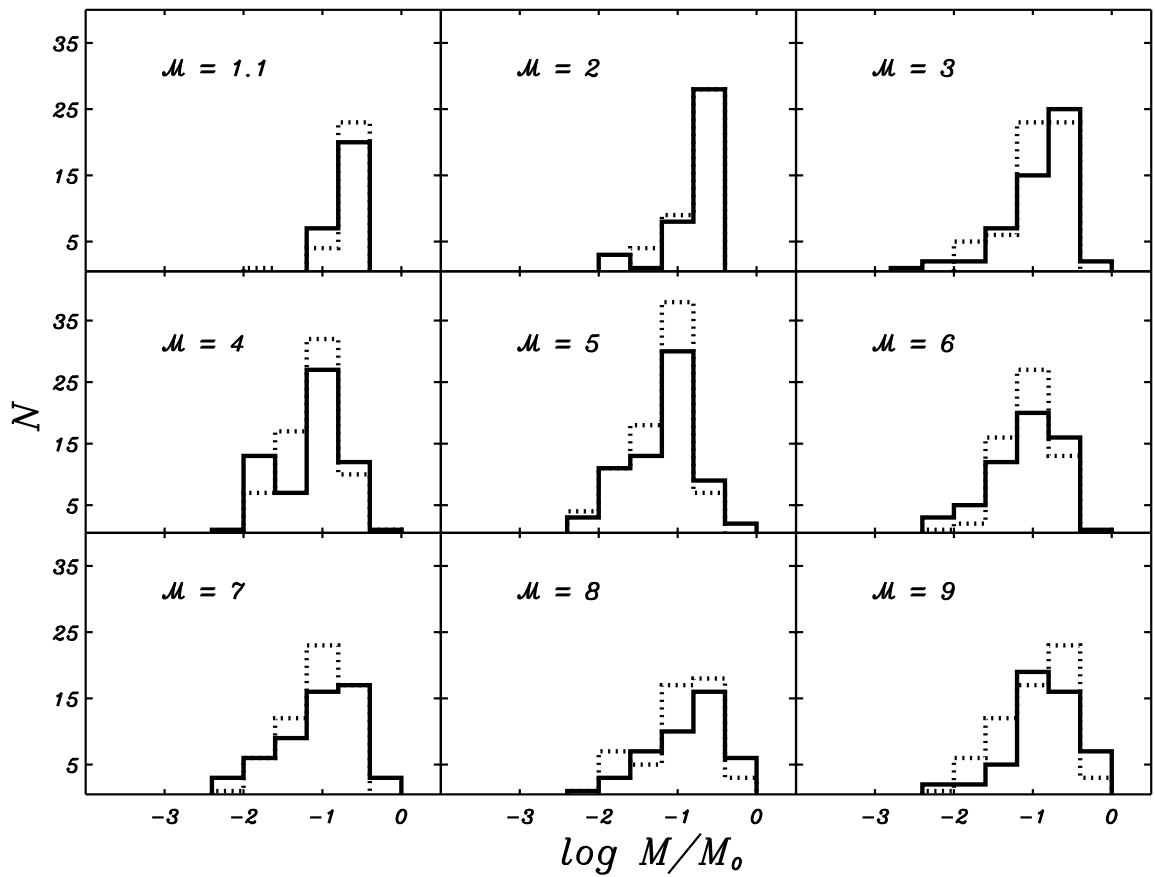}
}
\caption{Same as in Figure \ref{fig:hist_comp_l}, except for bound
GRID-cores (i.e mass is $M_\mathrm{3D,th}$ and
$M_\mathrm{2D,bs,th}$). When the condition $E_{\rm th}+E_g <0$ is
applied, most of the low mass cores are eliminated, for every Mach
number. The 2D bound GRID-cores have almost the same mass distribution
as 3D bound GRID-cores.}
\label{fig:hist_comp_l_t}
\end{figure}

Figure \ref{fig:hist_comp_l} shows histograms for the distributions of
$M_\mathrm{2D,bs}$ and $M_\mathrm{3D}$ (all GRID-cores) for each
$\mathcal{M}$, while Figure \ref{fig:hist_comp_l_t} shows the
histograms of $M_\mathrm{2D,bs,th}$ and $M_\mathrm{3D,th}$ (bound
GRID-cores), both for low perturbation amplitudes. The distributions
of $M_\mathrm{2D,bs}$ and $M_\mathrm{3D}$ are quite similar for all
$\mathcal{M}$, except slightly more low mass cores are identified for
2D at large $\mathcal{M}$. When the thermal - gravitational energy
condition is included in defining cores, the low-mass end of the
distribution is removed; in Fig. \ref{fig:hist_comp_l_t}, the 2D bound
GRID-cores have almost exactly the same distributions as 3D bound
GRID-cores.

\begin{figure}
\centerline{
\includegraphics[width=0.8\textwidth]{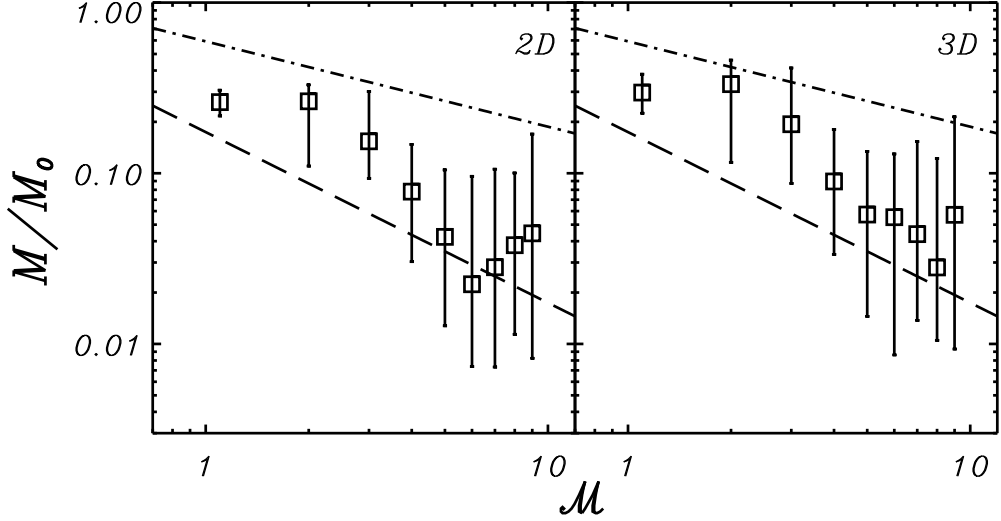}
}
\caption{ Median GRID-core mass $M$ versus Mach number $\mathcal{M}$
of the inflow. The left panel is for 2D GRID-cores
($M_\mathrm{2D,bs}$) and the right panel for 3D GRID-cores
($M_\mathrm{3D}$).  Vertical bars indicate quartiles of the
distribution. Also shown is the expected mass dependence for early
gravitational fragmentation given by equation (\ref{m_mgrn0}) (with
$M\propto \mathcal{M}^{-1/2}$, dot-dashed), and late gravitational
fragmentation given by equation (\ref{m_crit_sg1}) (with $M \propto
\mathcal{M}^{-1}$, dashed). The critical Bonnor-Ebert mass at the
post-shock density (see eq. \ref{m_be_ps}) is similar to the
late-stage prediction ($M \propto \mathcal{M}^{-1}$, dashed). The
relation between median core mass and $\mathcal{M}$ is quite similar
for 2D and 3D cores. Core mass declines with increasing Mach number
$\mathcal{M}$, lying between the $M\propto \mathcal{M}^{-1/2}$ (early
stage) and $M \propto \mathcal{M}^{-1}$ (late stage) fragmentation
predictions.  }
\label{fig:mass_comp_l}
\end{figure}

\begin{figure}[ht]
\centerline{
\includegraphics[width=0.8\textwidth]{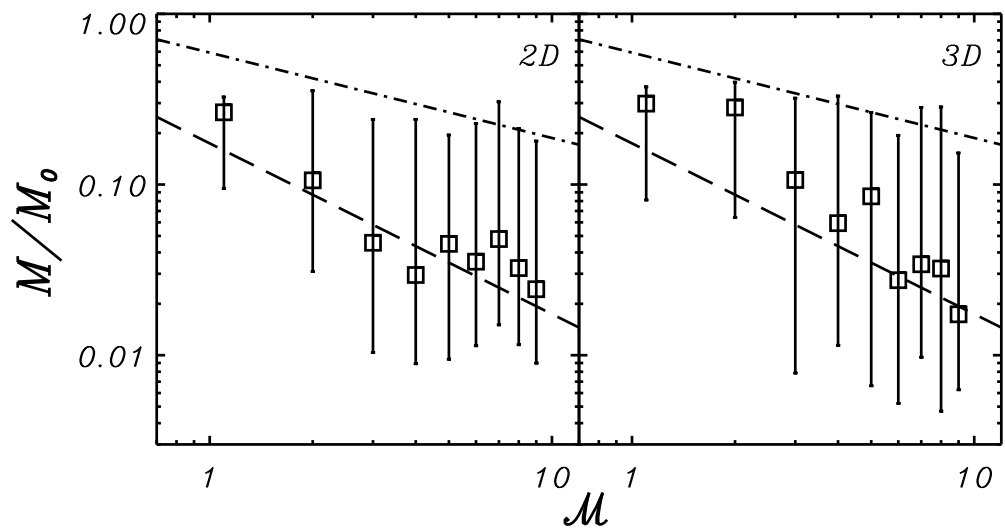}
}
\caption{ Median GRID-core mass $M_\mathrm{2D,bs}$ and
$M_\mathrm{3D}$, as shown in Figure \ref{fig:mass_comp_l}, but for
high amplitude initial perturbations. The median masses are slightly
smaller than for low amplitude initial perturbations, but follow a
similar trend.  }
\label{fig:mass_comp_h}
\end{figure}

Figure \ref{fig:mass_comp_l} (all GRID-cores) and Figure
\ref{fig:mass_comp_l_t} (bound GRID-cores) show the median core mass
(squares in figures) versus $\mathcal{M}$ from Figure
\ref{fig:hist_comp_l} and \ref{fig:hist_comp_l_t}, respectively. (We
do not measure the peak because some of the histograms are irregular.)
Figure \ref{fig:mass_comp_h} (all GRID-cores) and 
Figure \ref{fig:mass_comp_h_t}
(bound GRID-cores) show the same median mass -- $\mathcal{M}$ relation
for high amplitude initial perturbations. The breadth of the
distributions at each $\mathcal{M}$ is indicated by vertical bars: the
lower bar is the difference between the median and the first quartile,
and the higher bar is the difference between the third quartile and
the median.  In Fig. \ref{fig:mass_comp_l}, \ref{fig:mass_comp_l_t}
and Fig. \ref{fig:mass_comp_h}, \ref{fig:mass_comp_h_t}, we overlay
lines showing the predicted critical mass at late stages
(eq. \ref{m_crit_sg1} or \ref{m_crit_sg3}, dashed line with $M \propto
\mathcal{M}^{-1}$), and the prediction for the mass that has grown the
most at early time (eq. \ref{m_mgrn0} or \ref{m_mgrn2}, dot-dashed
with $M \propto \mathcal{M}^{-1/2}$). The post-shock Bonnor-Ebert mass
($M \propto \mathcal{M}^{-1}$ from eq. \ref{m_be_ps}) is similar to
the late-stage critical mass.

As the Mach number increases, the post-shock density $\rho \approx
\rho_0 \mathcal{M}^2$ is higher. This lowers the Jeans length (as well
as the Jeans mass and Bonnor-Ebert mass), permitting smaller (but
denser) cores to form at high $\mathcal{M}$ compared to low
$\mathcal{M}$.  However, high mass cores can still form at high
$\mathcal{M}$, as is evident in Figure \ref{fig:hist_comp_l} and
\ref{fig:hist_comp_l_t} and the quartiles shown in Figures
\ref{fig:mass_comp_l} -- \ref{fig:mass_comp_h_t}: at high
$\mathcal{M}$, the histograms extend to low mass, but the high mass
part of the distribution is still present.  This is consistent with
the expectation that any scale above the critical scale can grow more
nonlinear due to self-gravity (see eqs. \ref{disp_rl} -
\ref{crit_m}).

Based on Figures \ref{fig:mass_comp_l} -- \ref{fig:mass_comp_h_t}, we
also note that the median mass versus $\mathcal{M}$ relations are
quite similar whether cores are identified with the 2D or 3D
gravitational potential. This is true for low or high amplitude
perturbations, for both all GRID-cores and bound GRID-cores. This
evidently shows that 2D cores have similar statistical properties to
the 3D cores. Since the GRID algorithm is easy to implement for
observational data, it appears to be a promising method for finding
cores.\footnote{An IDL implementation of our GRID-core algorithm for
use with observed data (FITS files containing surface density maps) is
available from the authors.}

Median masses for GRID-cores decline with increasing Mach number for
both low and high amplitude perturbations (see
Figs. \ref{fig:mass_comp_l}, \ref{fig:mass_comp_h}).  These median
masses generally lie above the values predicted from equations
(\ref{m_be_ps}), (\ref{crit_m_min_nsg}) and (\ref{m_crit_sg1})
($M\propto \mathcal{M}^{-1}$) at late stages and below the values
predicted from equation (\ref{m_mgrn0}) ($M\propto
\mathcal{M}^{-1/2}$) at early stages. The median GRID-core masses for
high amplitude perturbations are slightly smaller than those for low
amplitude perturbations, and the range of core masses for a given Mach
number are larger. This reflects the fact that the percentage of small
cores is higher when the perturbation amplitudes are higher.  
GRID-cores are identified based on the gravitational potential, 
and this potential reflects density structure, which arises
from both turbulent and gravitational processes.  Even without
gravity, smaller scale masses would be expected in the
higher-$\mathcal{M}$ models because of their high turbulent
amplitudes.  For our simulations, the input perturbation amplitude at
scale $l$ is $\delta v_\mathrm{1D} (l) = (l/L_J)^{1/2}
(\mathcal{M}/3)^{1/2}\,c_s$ at $100\%$ amplitude of perturbation
(cf. eq. \ref{pert_ampl}). Structures at scales $l$ for which
turbulent perturbations are supersonic will, even in the absence of
gravity, be more prominent than those at smaller scale.  For our
adopted scaling of input perturbations with $\cal M$, the sonic
scale varies as $l_\mathrm{sonic} \propto L_J/\mathcal{M}$, so that
the mass at the sonic scale varies $\propto \Sigma(t)\,
l_\mathrm{sonic}^2$. With $\Sigma(t) \propto \mathcal{M}
t_\mathrm{coll}$ and $t_\mathrm{coll} \propto \mathcal{M}^{-1/2}$ (see
eq. \ref{t_mgrn} and below), this predicts $M_\mathrm{sonic} \propto
\mathcal{M}^{-3/2}$.  For later time $t \sim t_J$ (comparable to the
flow crossing time for a cloud with $\alpha_{\rm vir} =$ 1 -- 2),
$M_{\rm sonic} \propto \mathcal{M}^{-1}$.  Thus, the sonic mass scale,
and hence the mass scale of nonlinear structures induced purely by
turbulence, is expected to decline with increasing $\mathcal{M}$.

For bound GRID-cores, the median mass vs. $\mathcal{M}$ decreases and
then increases, for low amplitude perturbations
(Fig. \ref{fig:mass_comp_l_t}), and is nearly flat for high amplitude
perturbations (Fig. \ref{fig:mass_comp_h_t}).  The high median mass at
high $\mathcal{M}$ for bound GRID-cores may be due to a combination of
effects, including numerical resolution and nonlinearity.  The
characteristic scale for self-gravitating perturbations decreases with
increasing Mach number (either as $r \propto \mathcal{M}^{-1/2}$ for
the most-grown core or $r \propto \mathcal{M}^{-1}$ for critical
perturbations; see Section 2). At high $\mathcal{M}$, this may
approach or fall below the minimum scale $r_\mathrm{min} =4 $ zones $=
0.016L_J$ that we require for the GRID-core radius to be well
resolved. Since the post-shock density is $\propto \mathcal{M}^2$, the
GRID-core mass would then increase at least $\propto \mathcal{M}^2
r_\mathrm{min}^3$ at sufficiently high $\mathcal{M}$.  In addition,
larger-scale, higher-mass regions initially have higher amplitude
perturbations than smaller-scale regions, because of the input power
spectrum with $\delta v \propto l^{1/2}$. If this initial ``head
start'' allows the larger, more massive cores to become highly
nonlinear before more rapidly-growing smaller-scale cores, the more
massive cores will collapse (halting the simulation) before the
lower-mass cores become strongly concentrated (with $E_{\rm th} <
|E_g|$) internally.  With implementation of sink particles such that
the simulations need not to be halted when the most evolved core
collapses, and $|E_g|$ can grow for low-mass cores, it will be
possible to test whether the median mass of bound cores decreases with
increasing $\mathcal{M}$, similar to Figs.  \ref{fig:mass_comp_l} and
\ref{fig:mass_comp_h}.

\begin{figure}[ht]
\centerline{
\includegraphics[width=0.8\textwidth]{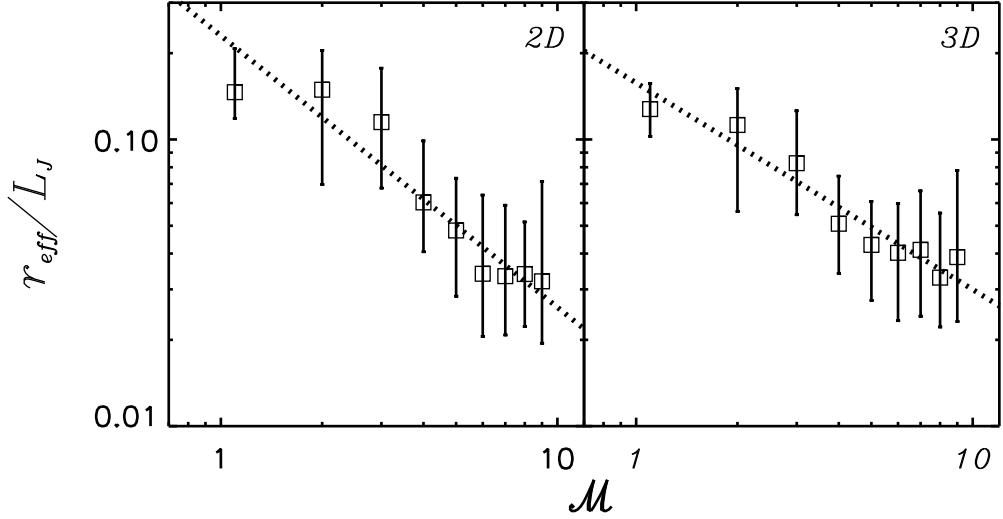}
}
\caption{Median GRID-core radius versus Mach number $\mathcal{M}$ for
low amplitude initial perturbations. Core sizes are defined using the
largest closed contours of the gravitational potential in 2D
($\Phi_{\rm 2D}$, left) and 3D ($\Phi$, right). Vertical bars indicate
quartiles of the distribution. The dotted lines are power-law fits:
$r_\mathrm{eff,2D,bs}/L_J = 0.23^{0.29}_{0.18}
\mathcal{M}^{-0.95\pm0.13}$ and $r_\mathrm{eff,3D}/L_J =
0.16^{0.18}_{0.14} \mathcal{M}^{-0.72\pm0.07}$.}
\label{fig:radius_comp_l}
\end{figure}

\begin{figure}[ht]
\centerline{
\includegraphics[width=0.8\textwidth]{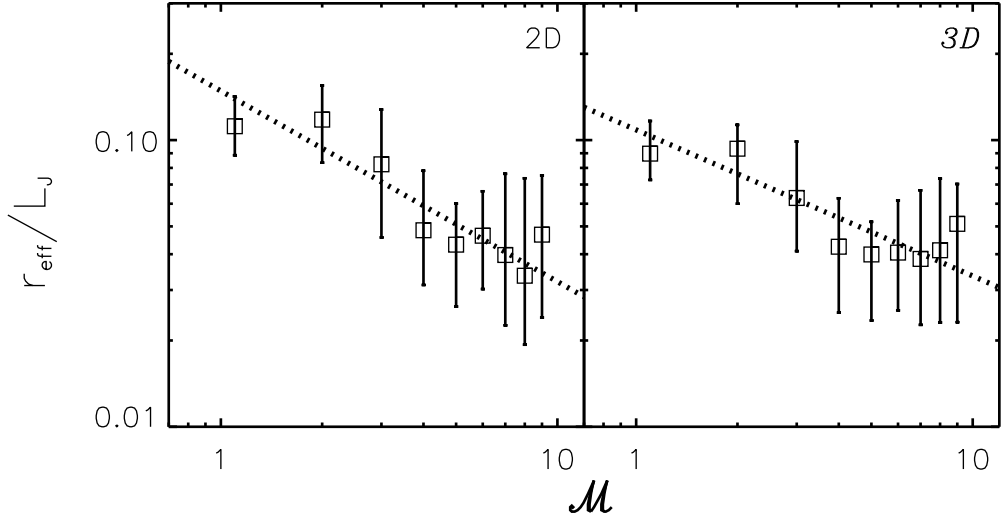}
}
\caption{Same as in Figure \ref{fig:radius_comp_l} but for bound
GRID-cores ($E_{\rm th}+ E_g <0$). The power-law fits are
$r_\mathrm{eff,2D,bs,th}/L_J = 0.15^{0.18}_{0.12}
\mathcal{M}^{-0.67\pm0.10}$ and $r_\mathrm{eff,3D,th}/L_J =
0.11^{0.12}_{0.10} \mathcal{M}^{-0.61\pm0.08}$}.
\label{fig:radius_comp_l_t}
\end{figure}

Figure \ref{fig:radius_comp_l} shows the GRID-core radii (as defined
in Section 3) versus Mach number, and Figure \ref{fig:radius_comp_l_t}
shows the bound GRID-core radii versus Mach number; these are for
cases with low amplitude initial perturbations. Overall, the median
radii for all GRID-cores and bound GRID-cores decrease towards higher
$\mathcal{M}$.  This is consistent with expectations: high Mach number
yields high post-shock density, and hence a smaller Jeans length; in
addition, the higher amplitude of input turbulence at higher
$\mathcal{M}$ makes the sonic scale smaller. The prediction for core
radius based on turbulence alone would be the sonic scale from
Equation (\ref{pert_ampl}): $r_\mathrm{eff} \propto l_\mathrm{sonic}
\propto L_J/\mathcal{M}$. The first core to collapse is predicted to
have $\lambda_m \propto \mathcal{M}^{-1/2}$ from equation
(\ref{lambda_mgrn}). For late-time fragmentation, the relevant scale
is the Jeans length in post-shock gas, which varies $\propto
\mathcal{M}^{-1}$. For GRID-cores, the slopes are between these
values, equal to $-0.95\pm0.13$ for $r_\mathrm{eff,2D,bs}$ and
$-0.72\pm0.07$ for $r_\mathrm{eff,3D}$, for low amplitude initial
perturbations.  For bound GRID-cores, the power-law fit for median
radius as a function of Mach number gives slope $-0.67\pm0.10$ and
$-0.61\pm0.08$ for 2D and 3D respectively.  These are comparable to
the result $\lambda_m \propto \mathcal{M}^{-1/2}$ from Equation
(\ref{lambda_mgrn}). Although the overall slopes are close to $-0.5$,
we note that the relation flattens at $\mathcal{M} \gtrsim 5$,
possibly due to our requirement that the effective radius must exceed
4 zones, or because the initial power spectrum favors larger cores.

Figure \ref{fig:time_comp} shows the median collapse time of the most
evolved core $\mathrm{vs.}$ Mach number, for both low and high
amplitude initial perturbations. They both follow power laws close to
$t_\mathrm{coll} \propto \mathcal{M}^{-1/2}$, consistent with the time
scale (see eq. \ref{t_mgrn}) predicted for growth of self-gravitating
modes up to a given amplification $\Gamma_\mathrm{max}$. The
coefficients for low amplitude initial perturbations and high
amplitude initial perturbations are 0.69 and 0.51, respectively,
compared to 0.34 from equation (\ref{t_mgrn}) taking
$\Gamma_\mathrm{max}=1$. With high amplitude initial perturbations,
cores collapse earlier because the seed perturbations need not grow as
much. Note that the naive expectation based on the Jeans time, taking
$\rho_\mathrm{post-shock} \propto \mathcal{M}^{-2}$, would yield a
steeper dependence $t \propto \rho_\mathrm{post-shock}^{-1/2} \propto
\mathcal{M}^{-1}$.  Based on Fig. \ref{fig:time_comp}, it is evident
that the first cores in higher $\mathcal{M}$ cases collapse when the
layer as a whole is only barely self-gravitating ($t_\mathrm{coll}/t_0
\sim 0.2-0.3$, compared to $t_\mathrm{sg} \approx 0.22 t_0$ from
eq. \ref{t_sg}), whereas the layer is more strongly self-gravitating
at the first collapse for low-$\mathcal{M}$ cases.

The shape of a core can be characterized by the eigenvalues of the
moment of inertia tensor $I_{ij} \equiv \int \rho x_i x_j d^3
\mathbf{x}$ \citep[e.g.][]{gamm03,naka08}. Let $a, b$ and $c$ be the
lengths of the principal axes and $a \ge b \ge c$. Then a prolate core
has $b/a = c/a$, and an oblate core has $b/a =1$. We have computed the
moment of inertia and aspect ratios for all the cores identified in
our simulations. For example, the aspect ratios of the most evolved
cores shown in Figures \ref{fig:comp_mach} and
\ref{fig:core_structure_2d} are $b/a = 0.39, c/a = 0.25$ for the
$\mathcal{M} = 1.1$ model and $b/a = 0.28, c/a = 0.25$ for the
$\mathcal{M} =5$ model.  They are both (approximately) prolate
according to the classification of \citet{gamm03}.

\begin{figure}[ht]
\centerline{
\includegraphics[width=0.7\textwidth,angle=90]{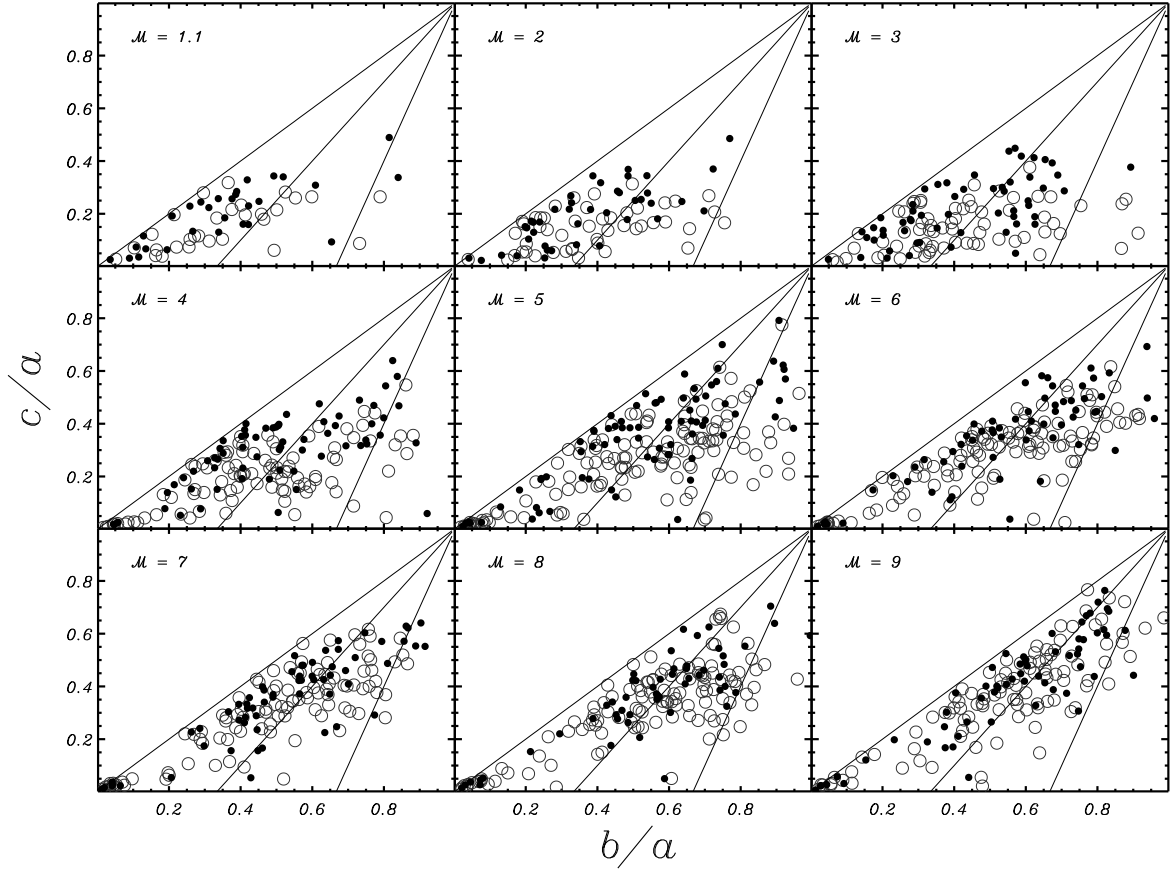}
}
\caption{Distribution of three-dimensional core aspect ratio for each
Mach number for low amplitude initial perturbations. Cores lying on
$c/a=b/a$ are formally prolate and along $b/a =1$ are formally oblate.
We subdivide (see diagonal lines) and classify as follows:
approximately prolate (between $c/a=1$ and $c/a=1.5b/a-0.5$), triaxial
(between $c/a=1.5b/a-0.5$ and $c/a=3b/a-2$) and approximately oblate
(between $c/a=3b/a-2$ and $b/a=1$). Open circles are GRID-cores
defined by the gravitational potential contours alone. Dots are bound
GRID-cores, with the additional requirement $E_{\rm th}+ E_g <0$.}
\label{fig:core_shape_l}
\end{figure}

\begin{figure}[ht]
\centerline{
\includegraphics[width=0.7\textwidth,angle=90]{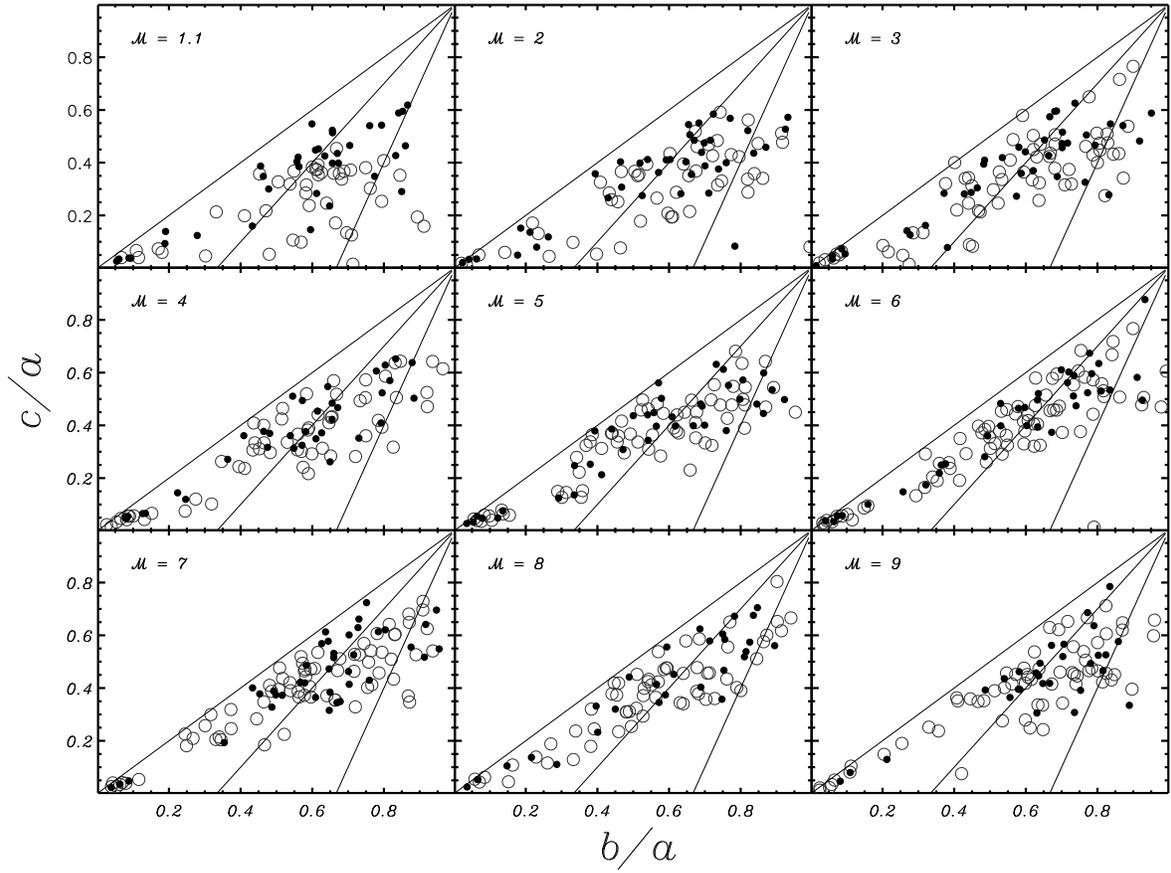}
}
\caption{Same as Figure \ref{fig:core_shape_l} but for high amplitude initial
perturbations.}
\label{fig:core_shape_h}
\end{figure}

Figure \ref{fig:core_shape_l} and Figure \ref{fig:core_shape_h} show
the distribution of core aspect ratios for each $\mathcal{M}$ for low
and high amplitude initial perturbations respectively.  Open circles
represent GRID-cores, and dots represent bound GRID-cores. These
distributions show a number of interesting features and trends. First,
only a small portion of cores are oblate for each $\mathcal{M}$, for
both low and high amplitude perturbations. Second, more oblate-like
cores appear when the initial perturbation amplitudes are higher. For
low amplitude perturbations, at $\mathcal{M} = 1.1$ and $2$, $c/a$ and
$b/a$ are mostly $\le0.5$, i.e. approximately prolate. But at larger
$\mathcal{M}$ for low amplitude initial perturbations, and all
$\mathcal{M}$ for high amplitude perturbations, there are many cores
in the triaxial and oblate regions.  Also, large and massive cores
tend to be more prolate. For low amplitude perturbations, at
$\mathcal{M}=1.1$, almost all the cores formed are prolate and no
small cores form (compared to high Mach number cases). The reason that
the distribution is more oblate for higher amplitude perturbation
(large $\mathcal{M}$ for low amplitude initial perturbations, and all
$\mathcal{M}$ for high amplitude initial perturbations) is that more
of the cores are at earlier stages of evolution. Figure
\ref{fig:comp_mach} shows development of cores for $\mathcal{M} = 1.1,
5$ and $8$.  As is particularly clear for the stages shown in the
$\mathcal{M}=1.1$ model, structures are more oblate during the
core-building stage than during the collapse stage.  Cores evolve to
become prolate when they collapse because the collapse happens first
in the directions perpendicular to the larger scale filaments. For
$\mathcal{M}=1.1, 2$ models with low amplitude perturbations, only
large cores form and they have evolved to the collapse stage and
become prolate. Models with higher amplitude perturbations have a
greater percentage of small cores that have not yet collapsed.

We can also examine the relationship between core structure and
kinematics in our simulations.  Figure \ref{fig:v_los} shows the
projected density field, velocity field and the velocity dispersion
field along the line-of-sight for the $\mathcal{M} = 5$ model shown in
Fig. \ref{fig:corefind_comp_2d3d}.  We ``view'' the simulation at
angles $0^\mathrm{\circ}, 30^\mathrm{\circ}$ and $60^\mathrm{\circ}$
with respect to the $z$ axis, tilting toward the $x$-axis.  The white
contours mark the regions identified as GRID-cores, and the orange
contours mark the bound GRID-cores. The projected density field is
smeared as the tilt angle $\theta$ increases. Since
$\left<v_\mathrm{los}\right> =
\left<v_x\right>\mathrm{sin}(\theta)+
\left<v_z\right>\mathrm{cos}(\theta)$, with
$\left<v_z\right>=0$ and the contribution from $\left<v_x\right>$ 
small at $\theta$ small, no obvious pattern is seen for
$\left<v_\mathrm{los}\right>$ at $\theta =0^\mathrm{\circ}$ and
$30^\mathrm{\circ}$.  At $\theta = 60^{\circ}$, when the
$\left<v_x\right>$ contribution becomes larger, converging flow
patterns similar to those seen in Fig. \ref{fig:vx_vy_proj} become
apparent, especially surrounding the diagonal line of small cores. As
previously discussed, converging flows in the $x$-$y$ plane create
this high density filament, which then fragments into small cores.

As Figure \ref{fig:v_los} shows, the dispersions of the line-of-sight
velocity of high density regions are generally subsonic, and are even
smaller in the cores. Velocity dispersions are low in high-density
regions for two reasons.  First, if filaments lie between supersonic
converging flows in the $x$-$y$ plane, then post-shock velocities
within the filaments will be subsonic. Second, weighting by density
picks out regions that are physically small along the
line-of-sight. The increase of linewidth with size means that if a
region is smaller than its surroundings along the line-of-sight, then
the linewidth will be smaller than that of its surroundings. Thus,
from a combination of low post-shock velocities (in the $x$-$y$
plane), and spatially-limited scale (in the $z$ direction),
$\sigma_\mathrm{los}$ is low in filaments and lower in cores, as seen
in Fig. \ref{fig:v_los}.

\section{Summary and discussion }
Stars form in GMCs pervaded by supersonic turbulence, and core
formation theory must take these supersonic turbulent flows into
account. In this work, we explore the physics of core formation in a
dynamic environment, focusing on post-shock layers generated by
collisions of supersonic flows. The framework we adopt --
three-dimensional planar converging flows containing multi-scale
turbulence -- enables us to analyze the internal structure and
kinematics of cores, and to investigate the relation between core
properties and the inflow Mach number $\mathcal{M}$. We consider a
range $\mathcal{M} = $ 1.1 -- 9, and conduct 180 simulations with
different realizations of the initial turbulent power spectrum, in
order to obtain a sizable statistical sample. In addition to core
masses and sizes, we measure aspect ratios. To define cores, we
introduce a new method based on the gravitational potential, and
compare properties of cores identified using $\Phi$ (from the volume
density) and $\Phi_\mathrm{2D}$ (from the plane-of sky projected
surface density).

Unlike previous studies of core evolution that begin with pre-existing
cores, the present models include formation stages. Our initial
density is uniform everywhere, and cores grow, via self-gravity, from
turbulence-induced perturbations within the post-shock layer; when the
Mach number is high, initial growth of density perturbations is aided
by shock-driven hydrodynamic instabilities.  Based on a set of
spherically-symmetric numerical simulations, \citet{gong09} proposed
four stages for core evolution in dynamic environments: core building,
core collapse, envelope infall, and late accretion. The key features
during core building and collapse described in \citet{gong09} are
verified here, for more realistic geometry.  As the supersonic flows
converge in a plane, two reversed shocks propagate outwards.  With its
high mean density, the stagnation layer between these two shock fronts
becomes an incubator for self-gravitating cores. When these cores
become sufficient stratified, they collapse. We halt the simulations
at the instant of singularity formation in the most evolved core,
because the time step becomes very short.

Based on the analysis of our simulations, our chief conclusions are as follows:

1. Cores with realistic properties are able to form in post-shock
dense layers within turbulent GMCs. Core building to become
supercritical takes $\sim$ 10 times as long as the subsequent ``outside-in''
collapse stage, which lasts a few $\times 10^5 \yr$. The duration of
the supercritical stage is consistent with observations of prestellar
core lifetimes \citep{ward07, enoc08, evan09}.

2. At the time of singularity formation, the radial density profile
within cores approaches the Larson-Penston asymptotic solution
$\rho = 8.86 c_s^2/(4 \pi G r^2)$ and the velocity approaches
the Larson-Penston limit $-3.28 c_s$. This is consistent with previous
studies of spherical core collapse (see Section 1 for
references). \citet{till04} also found that $\rho \propto r^{-2}$ in
their most massive cores, for turbulent simulations.  As in
\citet{gong09}, we therefore conclude that the Larson-Penston
asymptotic solution is an ``attractor'' for core collapse, no matter
how the collapse is initiated.

\begin{figure}[ht]
\centerline{
\includegraphics[width=0.65\textwidth,angle=90]{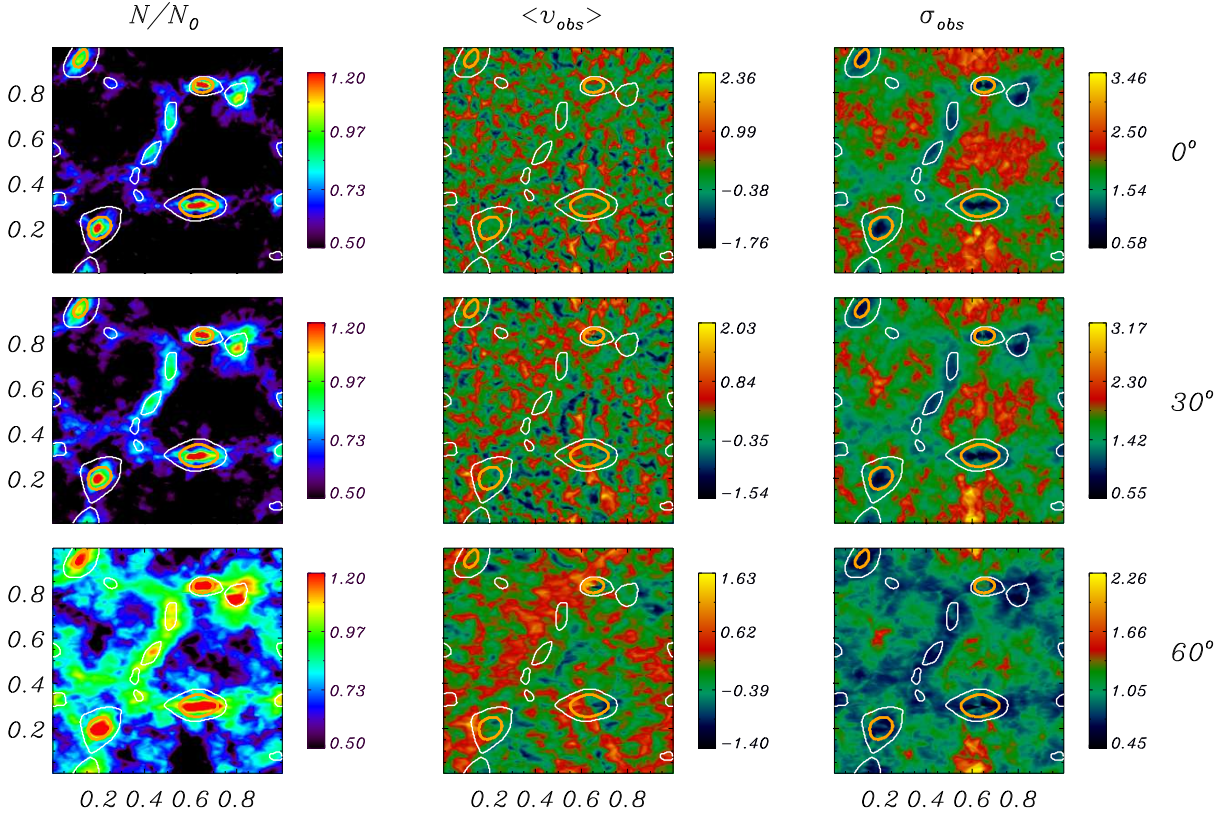}
}
\caption{Observations of one of the $\mathcal{M}=5$ models shown in
Fig. \ref{fig:corefind_comp_2d3d} from different angles. The first
column shows the surface density (color scale $\mathrm{log}
\Sigma/\Sigma_0$); the second column shows the line-of-sight velocity
and the third column shows the dispersion of the line-of-sight
velocity (linear color scale, in units of $c_s$).  The three rows from
top to bottom show the observed fields for $\theta_\mathrm{tilt} =
0^{\circ}, 30^{\circ}$ and $60^{\circ}$ respectively. The white curves
are the GRID-cores, and the orange curves are the bound
GRID-cores. Note that core regions have low internal velocity
dispersions.}
\label{fig:v_los}
\end{figure}

3.  Prior to collapse, the velocities within dense cores remain
subsonic, in spite of the highly-supersonic flows that create
them. This is true both for the ordered inflow, and for the mean
internal velocity dispersion.  This result is consistent with
observations that most cores have subsonic non-thermal velocity
dispersions \citep{myer83,good98,case02,tafa04,
  kirk07,andr07,lada08}. The velocity dispersion can increase quite sharply at
the edge of the core in our models (see Fig. \ref{fig:v_los}),
intriguingly similar to a sharp transition seen in $\mathrm{NH}_3$
observations by \citet{pine10} for the B5 core in Perseus.  From some
orientations, velocity dispersions in filaments containing cores may
also be lower than in the surrounding gas (cf. Fig. \ref{fig:v_los}).

4. At sub-pc scales, turbulent velocity perturbations (whether super-
or subsonic) induce density perturbations that can grow strongly if
the density is high enough for self-gravity to be important. In
post-shock layers, turbulence and self-gravity collect gas into long,
thin filamentary structures at the same time as the highest density
regions within the filaments grow to become centrally-condensed
cores. These filamentary structures containing embedded cores are
similar to the structures in the Aquila rift and Polaris Flare clouds
observed by {\it Herschel} \citep{andr10,mens10}.

5. Using the gravitational potential to identify cores is advantageous
because it enables a core definition based on dynamical
principles. For numerical simulations, the gravitational potential may
be computed from the volume density (yielding $\Phi$) or from the
projected surface density (yielding $\Phi_\mathrm{2D}$). We show for
our models that cores defined using $\Phi$ and $\Phi_\mathrm{2D}$ are
nearly the same, both for GRID-cores (defined by the largest closed
potential isosurfaces) and bound GRID-cores (which additionally
require $E_{\rm th} + E_g <0$). Since $\Phi_\mathrm{2D}$ can be
computed for observed clouds, using potential contours offers a
promising new core identification method for application to
high-resolution molecular cloud maps.  IDL code implementing our
GRID-core algorithm, suitable for application to observed data, is
available from the authors.

6. We find that the range of core masses that forms increases as the
Mach number $\mathcal{M}$ increases. Physically, this is because a
larger range of spatial scales has significant perturbations when the
turbulence amplitude is higher, and because the minimum mass to be
gravitationally unstable decreases as the density in the shocked layer
increases.  \citet{basu09} also found broader mass distributions when
the turbulent amplitude is increased.  At high Mach number, GRID-core
masses range between $\sim 10^{-3}$ -- $1 M_J$, corresponding to $\sim
0.05$ -- $50 \Msun$ for typical GMC conditions.

7. Analytical arguments (see Section 2) suggest that the first core to
collapse will have mass $M \propto \mathcal{M}^{-1/2}$, and that at
late times, the minimum mass core will vary as $M \propto
\mathcal{M}^{-1}$. Our numerical results for median core masses as a
function of $\mathcal{M}$ lie between these two relations. When the
core definition includes the condition that $E_{\rm th} + E_g < 0$,
the median mass increases at the largest Mach number. This may be due
to the nonlinear ``head start'' of massive cores, such that lower mass
cores have not yet become concentrated when the first core collapses
(and the simulation is stopped).

8. Analytical arguments (see Section 2) suggest that the effective
core radius will decline with increasing Mach number, with powers
between $r_\mathrm{eff} \propto \mathcal{M}^{-1/2}$ and
$r_\mathrm{eff} \propto \mathcal{M}^{-1}$. Our numerical results show
a decrease of $r_\mathrm{eff}$ with $\mathcal{M}$ in this range. For
bound GRID-cores ($E_{\rm th} + E_g < 0$), the relation is shallower
than for GRID-cores defined by gravitational potential alone.

\begin{figure}[ht]
\centerline{
\includegraphics[width=0.8\textwidth]{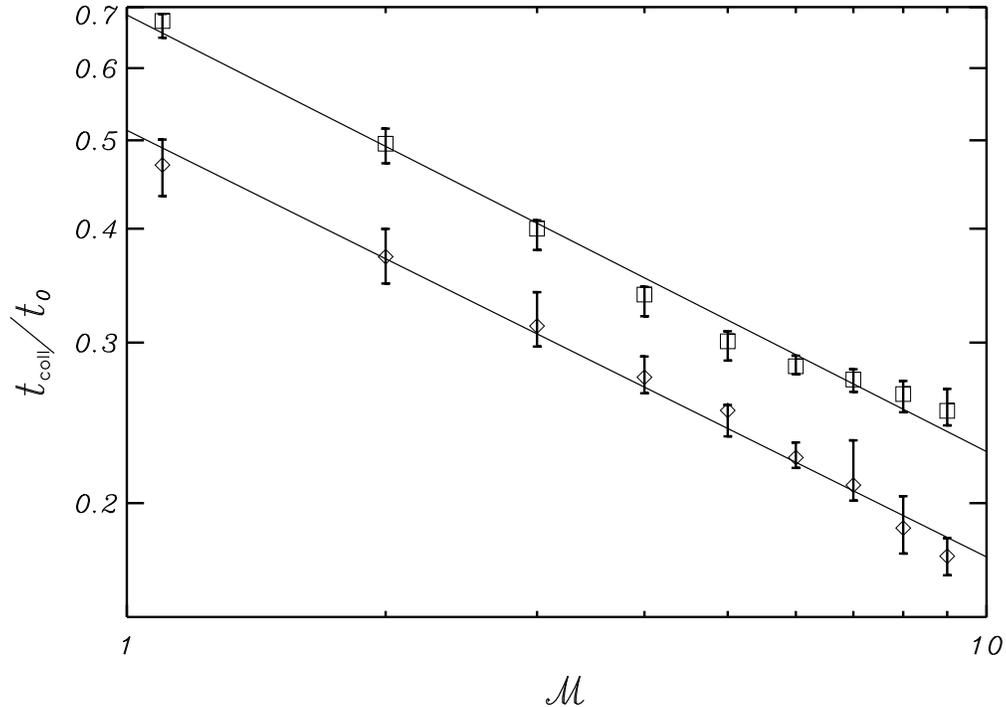}
}
\caption{Collapse time of the most evolved core, $t_{\mathrm{coll}}$,
versus inflow Mach number $\mathcal{M}$ for low amplitude (squares)
and high amplitude (diamonds) initial perturbations. Each value is the
median of $t_{\mathrm{coll}}$ for 20 simulations for each
$\mathcal{M}$. Vertical bars indicate quartiles of these 20 values of
$t_{\mathrm{coll}}$. The solid line least-squares fits are:
$t_\mathrm{coll}/t_0 = 0.69 \mathcal{M}^{-0.48}$ (low amplitude) and
$t_\mathrm{coll}/t_0 = 0.51 \mathcal{M}^{-0.47}$ (high amplitude).
The scaling is comparable to $t_\mathrm{coll} \propto
\mathcal{M}^{-0.5}$, as predicted by equation (\ref{t_mgrn}).  
The simulation time unit $t_0$, based on the mean inflow density, is
given in equation (\ref{tJ_def}).
}
\label{fig:time_comp}
\end{figure}

9. The time for the first core to collapse in our simulations depends
on Mach number, with $t_\mathrm{coll} \propto \mathcal{M}^{-1/2}$, and
a slightly smaller coefficient for high-amplitude initial
perturbations (see Fig. \ref{fig:time_comp}). This scaling is
consistent with analytic predictions for gravitational instability in
a shocked converging flow (see eq. \ref{t_mgrn}).  For high
$\mathcal{M}$, as is observed in GMCs, the first cores could collapse
within a few Myr of cloud formation. For high $\mathcal{M}$, the first
cores collapse when the shocked layer containing them is only barely
self-gravitating; this suggests that collections of stars can begin to
form individually before they collapse together to create a cluster.

10. A very small portion of cores are oblate, while most cores are
prolate or triaxial.  Large cores are preferentially prolate. The
triaxiality of most cores is consistent with previous results from
turbulent hydrodynamic and MHD simulations
\citep{gamm03,li04,naka08,offn08}.  We also find that core shapes
change as they evolve, from more oblate during early stages to more
prolate during collapse. For high initial perturbation amplitudes, the
distributions have a higher proportion of oblate cores because small
cores are less evolved (at the time the first core collapses),
compared to those in models with low initial perturbation amplitudes.

As noted above, the current models have provided evidence that the
masses of cores that form depend not just on the mean Jeans mass in a
cloud, but also on the cloud's level of internal turbulence at large
scales, $\sigma_v$. Equations (\ref{m_crit_sg1}) and
(\ref{m_crit_sg3}) suggest that at late times, the characteristic core
mass will follow $M_c \propto \sigma_v^{-1}\rho_0^{-1/2}T^2$, where 
$\rho_0$ is the mean density in the cloud. For the
current simulations, however, we halt at the instant when the most evolved
core collapses (because the time step becomes very short). This limits
the condensation of small cores; they are present, but not yet
strongly bound. In order to fully test the dependence of $M_c$ on
cloud parameters, it is necessary to implement sink particles
\citep[e.g.][]{krum04,fede10} so that the simulation can run until all
the ``eligible'' cores in the post-shock region have had the
opportunity to collapse. Including sink particles, as well as studying
shocked converging flows within larger turbulent clouds via
mesh-refined simulations, represent important avenues for future
research.

\acknowledgements
We are grateful to Lee Mundy and Alyssa Goodman for stimulating
conversations, and to the referee for a helpful report. This work was
supported by grants NNX09AG04G and NNX10AF60G from NASA.

\newpage

\newpage

\newpage

\newpage

\newpage

\newpage

\newpage

\newpage

\newpage

\newpage

\newpage

\newpage

\begin{figure}[ht]
\centerline{
\includegraphics[width=0.8\textwidth]{fig13.epsi}
}
\caption{Same as in Figure \ref{fig:mass_comp_l}, but for bound
GRID-cores ($E_{\rm th} + E_g <0$, i.e. $M$ is $M_\mathrm{2D,bs,th}$
or $M_\mathrm{3D,th}$).}
\label{fig:mass_comp_l_t}
\end{figure}

\begin{figure}[ht]
\centerline{
\includegraphics[width=0.8\textwidth]{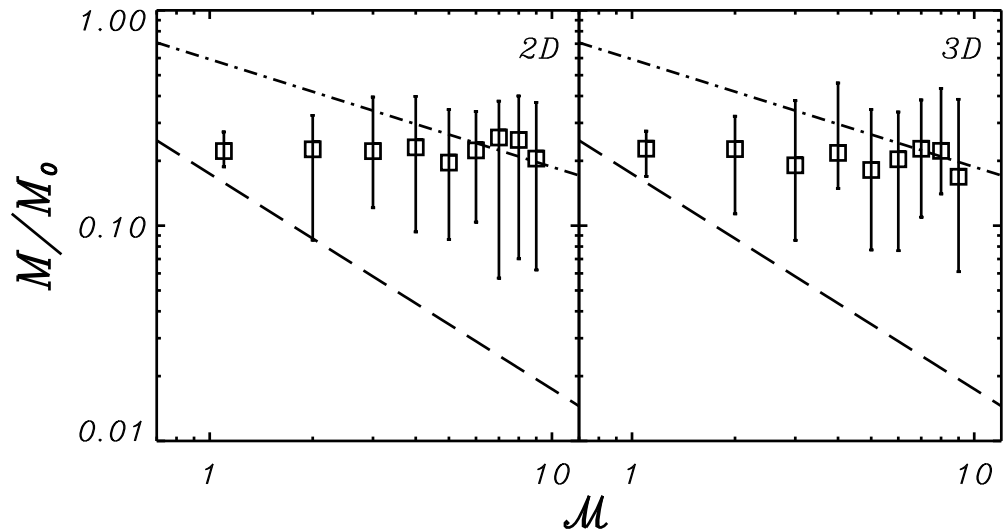}
}
\caption{Median bound GRID-core mass $M_{\rm 2D,bs,th}$ and $M_{\rm
3D,th}$ (i.e. $E_{\rm th} + E_g <0$) as in Figure
\ref{fig:mass_comp_l_t}, but for high amplitude initial perturbations.
}
\label{fig:mass_comp_h_t}
\end{figure}

\clearpage

\end{document}